\documentclass[aps,pra,showpacs,twocolumn,amsmath,amssymb]{revtex4}

\usepackage{graphics}
\usepackage{epsfig}

\newcommand{\ensm}[1]{\ensuremath{#1}}

\newcommand{\aiaj}[1]{\ensm{#1^{\dagger}_i#1_j}}
\renewcommand{\alph}{\ensm{\alpha}}

\newcommand{\avt}[1]{\ensm{\left\langle#1\right\rangle_T}}

\newcommand{\avtka}[1]{\ensm{\left\langle#1\right\rangle_{T,K}^{\alph}}}
\newcommand{\avtkb}[1]{\ensm{\left\langle#1\right\rangle_{T,K}^{\beta}}}
\newcommand{\Az}{\ensm{A(z)}}

\newcommand{\bra}[1]{\ensm{\langle #1|}}

\newcommand{\braai}{\ensm{\langle\alpha,i|}}

\newcommand{\cbra}[1]{\ensm{^c\langle #1|}}
\newcommand{\cket}[1]{\ensm{| #1 \rangle^c }}

\newcommand{\Deltaij}{\ensm{\Delta_{ij}}}

\newcommand{\dqab}{\ensm{\delta q_{\alph\beta}}}
\newcommand{\dqcd}{\ensm{\delta q_{\gamma\delta}}}

\newcommand{\fNn}{\ensm{f_N^n}}

\newcommand{\Heff}{\ensm{H_{eff}}}
\newcommand{\Hint}{\ensm{H_{int}}}
\newcommand{\HO}{\ensm{H_{0}}}

\newcommand{\infint}{\ensm{\int_{-\infty}^{\infty}}}

\newcommand{\J}{\ensm{J}}

\newcommand{\JsqoverV}{\ensm{\frac{J^2}{V}}}

\newcommand{\ket}[1]{\ensm{| #1 \rangle}}

\newcommand{\ketai}{\ensm{|\alpha,i\rangle}}
\newcommand{\ketaj}{\ensm{|\alpha,j\rangle}}

\newcommand{\ketbj}{\ensm{|\beta,j\rangle}}
\newcommand{\Kij}{\ensm{K_{ij}}}

\newcommand{\M}{\ensm{M}}
\newcommand{\Mi}{\ensm{M_{i}}}
\newcommand{\Mj}{\ensm{M_{j}}}

\newcommand{\Om}{\ensm{\Omega}}

\newcommand{\psum}{\ensm{\sum_{(i,j)}}}

\newcommand{\Qaa}{\ensm{Q_{\alph\alph}}}
\newcommand{\Qab}{\ensm{Q_{\alph\beta}}}

\newcommand{\qab}{\ensm{q_{\alph\beta}}}
\newcommand{\QEA}{\ensm{Q_{EA}}}

\newcommand{\rh}{\ensm{\rho}}
\newcommand{\rhokt}{\ensm{\rh_{T,K}}}
\newcommand{\rhoakt}{\ensm{\rh^{\alph}_{T,K}}}

\newcommand{\Sone}{\ensm{S_1}}
\newcommand{\Stwo}{\ensm{S_2}}
\newcommand{\sa}{\ensm{s^{\alph}}}
\newcommand{\sbb}{\ensm{s^{\beta}}}
\newcommand{\scc}{\ensm{s^{\gamma}}}
\newcommand{\sd}{\ensm{s^{\delta}}}
\newcommand{\sech}{\ensm{\mbox{sech}}}
\newcommand{\si}{\ensm{s_i}}
\newcommand{\sia}{\ensm{s_i^{\alph}}}
\newcommand{\sib}{\ensm{s_i^{\beta}}}
\newcommand{\sj}{\ensm{s_j}}

\newcommand{\Nnspintr}{\ensm{\sum_{\left\{\sia=\pm 1\right\}}}}
\newcommand{\nspintr}{\ensm{\sum_{\left\{s^{\alph}=\pm 1\right\}}}}

\newcommand{\suma}{\ensm{\sum_{\alph}}}

\newcommand{\sumalb}{\ensm{\sum_{\alph<\beta}}}
\newcommand{\sumpab}{\ensm{\sum_{(\alph,\beta)}}}
\newcommand{\sumi}{\ensm{\sum_{i}}}

\newcommand{\lrb}{\ensm{\left(}}
\newcommand{\rrb}{\ensm{\right)}}
\newcommand{\lsb}{\ensm{\left[}}
\newcommand{\rsb}{\ensm{\right]}}
\newcommand{\lgb}{\ensm{\left\{}}
\newcommand{\rgb}{\ensm{\right\}}}
\newcommand{\Palpha}{\ensm{P_{\alpha}}}

\newcommand{\Pbeta}{\ensm{P_{\beta}}}
\newcommand{\Qalphai}{\ensm{Q_{\alpha i}}}
\newcommand{\Qalphaj}{\ensm{Q_{\alpha j}}}

\newcommand{\V}{\ensm{V}}

\newcommand{\wakt}{\ensm{w^{\alph}_{T,K}}}

\begin{document}

\title{Disordered ultracold atomic gases in optical lattices: A case study of Fermi-Bose mixtures}
\author{V. Ahufinger$^{1,2}$,
L. Sanchez-Palencia$^{2,3}$,
 A. Kantian$^{2,4,5}$, A. Sanpera$^{2,6,}$\footnote{also Instituci\'o Catalana de Recerca i
Estudis Avan\c cats}, and
M. Lewenstein$^{2,7,*}$}

\affiliation{$^1$Grup d'\`Optica, Departament de F\'isica, Universitat Aut\`onoma de Barcelona, E-08193 Belaterra,
Barcelona, Spain\\
$^2$Institut f\"ur Theoretische Physik, Universit\"at Hannover, D-30167 Hannover, Germany \\
$^3$Laboratoire Charles Fabry, Institut d'Optique Th\'eorique et Appliqu\'ee, Universit\'e
Paris-Sud XI, F-91403 Orsay cedex, France \\
$^4$Institut f\"ur Quantenoptik und Quanteninformation
der \"Osterreichischen Akademie der Wissenschaften, A-6020 Innsbruck, Austria \\
$^5$Institut f\"ur Theoretische Physik, Universit\"at Innsbruck, A-6020 Innsbruck, Austria\\
$^6$Grup de F\'isica Te\`orica, Departament de F\'isica, Universitat Aut\`onoma de Barcelona, E-08193 Belaterra, Barcelona, Spain \\
$^7$Institut de Ci\`encies Fot\`oniques, E-08034 Barcelona, Spain}

\date{\today}

\begin{abstract}
We present a review of properties of ultracold atomic Fermi-Bose mixtures in inhomogeneous and random optical lattices.
In the strong interacting limit and at very low temperatures,
 fermions form, together with bosons or bosonic holes, {\it composite fermions}. Composite fermions behave
as a spinless interacting
 Fermi gas, and in the presence of local disorder they interact via random couplings and feel effective
random local potential. This opens a wide variety of possibilities of realizing various kinds of ultracold
quantum disordered systems.
In this paper we review these possibilities, discuss the accessible quantum disordered phases, and methods
for their detection. The discussed quantum phases include Fermi glasses, quantum spin glasses, "dirty" superfluids,
disordered metallic phases,
and phases involving quantum percolation.
\end{abstract}
\maketitle

%%%%%%%%%%%%%%%%%%%%%%%%%%%%%%%%%%%%%%%%%%%%%%%%%%%%%%%%%%%%%

\section{Introduction}

\label{sec:1}

\subsection{Disordered systems}
Since the discovery of the quantum localization phenomenon  by P.W. Anderson
in 1958 \cite{pwa}, disordered and frustrated systems have played a central
role in condensed matter physics. They have been involved in some of the most
challenging open questions concerning many body systems (cf. \cite{cusack,balian,chow}). Quenched (i.e.,
frozen on the typical time scale of the considered systems) disorder
determines the physics of various phenomena, from transport and conductivity,
through localization effects and metal-insulator transition (cf. \cite{anderson}),
 to spin glasses (cf. \cite{parisi,stein,sachdev}), neural networks
(cf. \cite{amit}), percolation \cite{percbook,spiperc}, high $T_c$ superconductivity
(cf. \cite{Auerbach}), or quantum chaos \cite{haake}. One of the reasons
why disordered systems are very hard to describe and simulate is related
to the fact that, usually, in order to characterize the system, one should calculate the relevant
physical quantities averaged over a particular realization of the
disorder. Analytical approaches require the averaging of, for instance, 
the free energy, which (being proportional to the logarithm of the partition
function in the canonical ensemble) is a very highly nonlinear function of the
disorder. Averaging requires then the use of special methods, such as the replica
trick (cf. \cite{parisi}), or supersymmetry method \cite{efetov}.
In numerical approaches this demands either simulating very large
samples to achieve ``self-averaging'',
or numerous repetitions of simulations of small samples. Obviously,
this difficulty is particularly important for quantum disordered systems.
Systems which are not disordered but frustrated (i.e., unable to fulfill simultaneously
all the constrains imposed by the Hamiltonian), lead very often to
similar difficulties, because quite often they are characterized
at low temperature by an enormously large number of  low energy excitations
(cf. \cite{lhuillier}). It is thus desirable to ask whether atomic, molecular physics and quantum optics may help to understand such systems.
In fact, very recently, it has been proposed
how to overcome the difficulty of quenched averaging
by encoding quantum mechanically in a superposition state of an auxiliary system, all possible realizations of the set of random parameters \cite{belen}. 
In this paper we propose a more direct
approach to the study of disorder: direct realization of various disordered models using cold atoms in optical lattices.

\subsection{Disordered ultracold atomic gases}

In the recent years there has been an enormous progress 
in the studies of ultracold weakly interacting \cite{lev}, as well as strongly
correlated atomic gases. In fact, present experimental techniques 
allow to design, realize and control in the laboratory various types of ultracold interacting
Bose or Fermi gases, as well as their mixtures.  Such ultracold gases can be
transfered to optical lattices and form a, practically perfect, realization of
various Hubbard models \cite{hubbards}. This observation, suggested in the seminal
theory paper by Jaksch {\it et al.} \cite{jaksch} in 1998, and confirmed
then by the seminal experiments of M. Greiner {\it et al.} \cite{bloch}, has
triggered an enormous interest in the studies of strongly correlated quantum
systems in the context of atomic and molecular lattice gases.

It became soon clear that one can introduce local disorder and/or frustration to such systems in a
controlled way using various experimentally feasible methods. Local quasi-disorder potentials may be
created by superimposing  superlattices incommensurable with the main one to the system. Although
strictly speaking such a superlattice is not disordered, its effects are very similar to those induced
by the genuine random potentials \cite{boseglass,keith,laurent}.
Controlled local truly random potentials can be created by placing a speckle pattern on the main 
lattice \cite{grynberg,dainty}. 
As shown in Ref. \cite{boseglass},  for a system of
strongly correlated bosons located in such a disordered lattice,
both methods should permit to achieve an Anderson-Bose glass \cite{fisher}, provided that the correlation
length of the disorder $L_{dis}$ is much smaller than the size of the system. Unfortunately,
it is difficult to have $L_{dis}$ smaller than few microns using speckles. Thus, $L_{dis}$  is typically larger
than the condensate healing length $l_{heal}=1/\sqrt{8\pi n a}$, where $n$ is the condenste density, and 
$a$ the atomic scattering length.  Due to this fact, 
i.e. due to the effects resulting from the nonlinear interactions, it is difficult
to achieve the Anderson localization regime with weakly interacting Bose-Einstein condensates (BECs)
\cite{inguscio,aspect}. We have shown, however, that quantum localization should be experimentally feasible
using the quasi-disorder created by several lasers with incommensurable frequencies \cite{arlt}.
Random local disorder appears also,
naturally, in magnetic microtraps and atom chips as a result of roughness of the underlying
surface (\cite{chiprough}, for theory see \cite{henkel,lukin}).

One can also create disorder using a second atomic species, by rapidly quenching it
from the superfluid to the localized Mott phase.
After such process, different lattice sites are populated by a random number of atoms of the second species,
which act effectively as random scatterers for the atoms of the first species \cite{privat1}. 
Last, but not least it is possible to use Feschbach resonances
in fluctuating or inhomogeneous magnetic fields in order to induce a novel type of disorder that corresponds to random, or at least
inhomogeneous nonlinear interaction couplings \cite{privat2}, (for theory in 1D systems
see \cite{diplom,santosdis}). It has been also been proposed \cite{anna}  
that tunneling induced interactions in systems with local disorder
results in controllable disorder on the level of next neighbor interactions. That opens a possible path for
the realization of quantum spin glasses \cite{anna}. As we have already mentioned, 
several experimental groups have already achieved
\cite{inguscio,aspect,arlt}, or are soon going to realize \cite{privat1,privat2} disordered potentials using these methods. 
It is worth mentioning here a very recent attempt to create controlled disorder using optical tweezers methods \cite{privat3}.

There are also several ways  to realize non-disordered, but frustrated systems with atomic lattice gases.
One is to create such gases
in ``exotic'' lattices, such as the kagom\'e lattice \cite{kagome}, another is to induce and control
the nature and range of interactions by adjusting the external optical potentials, such as, for instance, proposed in Ref.~\cite{demler}. Another
example of such situation
is provided for instance by  atomic gases in a 2-dimensional lattice interacting via dipole-dipole
interactions with dipole moment
polarized parallel to the lattice \cite{goral}.

Finally, there are also several ideas concerning the possibility of realizing various types of
complex systems using atomic lattice gases or  trapped ion chains. Particularly interesting are
here the possibilities of producing long range interactions (falling off as inverse of the square,
or cube of the distance) \cite{porras}, systems with several metastable energy minima, and last,
but not least systems in "designed" external magnetic \cite{hof}, or even non-abelian gauge fields \cite{oster}.

\subsection{Quantum information with disordered systems}

One important theoretical aspect that should be considered in this context deals with the role of
entanglement in quantum statistical physics in general (where it concerns quantum phase transitions,
entanglement correlation length and scaling \cite{entscal}), and characterization of various types of
distributed entanglement. This aspect is particularly interesting in theoretical and
experimental studies of disordered systems. The question which one is tempted to ask is, to what
extend one can realize quantum information processing in i) disordered systems,
ii) non-disordered systems with long range interactions, iii) non-disordered frustrated systems.

At the first sight, the answer to this question is negative. Disordered quantum information
processing sounds like {\it contradictio in adjecto}. But, one should not neglect the possible
advantages offered by the systems under investigation. First, such systems have typically a significant
number of (local) energy minima, as for instance happens in spin glasses. Such metastable states might
be employed to store information distributed over the whole system, as in neural network models.
The distributed storage implies redundancy similar to the one used in error correcting protocols \cite{protocols}.
Second, in the systems with long range interactions the stored information is usually robust:
metastable states have large basins of attraction thermodynamically, and destruction of a part
of the system does not destroy the metastable states (for the preliminary studies see Refs. \cite{briegel,us}).
Third, and perhaps the most important aspect for the present paper is that atomic ultracold gases offer a
unique opportunity to realize special purpose quantum computers (quantum simulators) to simulate
quantum disordered systems. The importance of the experimental realizations of such quantum simulators
will without doubts forward our understanding of quantum disordered systems enormously. In particular,
we can think about large scale quantum simulations of the  Hubbard model for spin 1/2 fermions with
disorder, which lies at the heart of the present-day-understanding of high $T_c$ superconductivity.
The impact of this possibility for physics and technology is hard to overestimate. Fourth,
very recently, several authors have used the ideas of  quantum information theory to develop novel
algorithms for  efficient simulations of quantum systems on classical computers \cite{vidal}.
Applications of these novel algorithms to disordered systems are highly desired.

\subsection{Fermi-Bose mixtures}

The present paper deals with the above formulated questions, which lie at the frontiers of
the modern theoretical physics, and concern not only atomic, molecular and optical (AMO) physics and quantum optics,
but also condensed matter physics, quantum  field theory, quantum statistical physics, and quantum information.
This interdisciplinary theme is one of the most hot current subjects of the physics of ultracold gases.
In particular, we present here the study of a specific example of disordered ultracold atomic gases:
Fermi-Bose (FB) mixtures in  optical lattices
in the presence of additional inhomogeneous and random potentials.

In the absence of disorder and in the
limit of strong atom-atom interactions such systems can be described in terms of
composite fermions consisting of a bare fermion, or a fermion paired with 1 boson (bosonic hole), or 2 bosons
(bosonic holes), etc. \cite{kagan}. The physics of Fermi-Bose mixtures in this regime has been studied by
us recently in a series of papers \cite{lewen,fb,fboptexp}; for contributions of other groups to the studies of
FB mixtures in traps and in optical lattices
see Ref. \cite{bulk} and for the studies of strongly correlated  FB mixtures in lattices see
Ref. \cite{other}, respectively. In particular, the validity of the effective Hamiltonian for fermionic
composites in 1D
was studied using exact diagonalization and the Density Matrix Renormalization Group method  in
Ref. \cite{mehring}. The effects of inhomogeneous trapping potential on FB lattice mixtures has been for the first
time discussed by Cramer {\it et al.}
\cite{neweisert}. The physics of disordered FB lattice mixtures was studied by us in Ref. \cite{anna},
which has essentially demonstrated that this systems may
serve as a paradigm fermionic system to study a variety of disordered phases and phenomena: from Fermi glass to quantum spin glass and quantum percolation.

\subsection{Plan of the paper}

The main goal of the present paper is to present the physics of the disordered FB lattice gas in more detail,
and in particular to investigate conditions
for obtaining various quantum phases  and
 quantum states of interest.

The paper is organized as follows. Section II
describes the ``zoology'' of disordered systems and disordered phases known from condensed matter physics.
We pay particular attention to the systems realizable with cold atoms on one side, and particularly
interesting from the other. This latter phrase means that we consider here the systems that concern
important open questions and challenges of the
physics of disordered systems. In this sense this section is thought as a list of such challenging open
questions that can be perhaps addressed by cold gases community. This section is thus directed to the
 cold gases experts, and is supposed to motivate and stimulate their interest in the physics of ultracold
disordered systems.

In section III, we introduce the composite fermions formalism, first discussing it for the case of homogeneous lattices, and then for disordered ones. We derive here the explicit formulae for the effective Hamiltonian, and for various
types of disorder. One of the results of this section concerns the generalizations of the results of Ref. \cite{anna}, that implies that {\it local} disorder
on the level of Fermi-Bose Hubbard model leads to randomness of the nearest neighbor tunneling and coupling coefficients for the  composite fermions. Obviously,
these tunneling and
coupling coefficients arise in effect of tunneling mediated interactions between the composites.

In section IV, we present our numerical results in the weak disorder limit,
 based on the time dependent Gutzwiller ansatz. These results concern mainly the physics
of composites, the realization of Fermi glass, and the transition from Fermi liquid to Fermi glass.

The results for the case of strong disorder, spin glasses, are discussed in section V. The problem of detection of the phenomena predicted in this paper is addressed in Section VI, whereas we conclude in section VII.

%%%%%%%%%%%%%%%%%%%%%%%%%%%%%%%%%%%%%%%%%%%%%%%%%%%%%%%%%%%%%
\section{Disordered systems: the old and new challenges for AMO physics}

In this chapter we present a list of problems and challenges of the physics
of disordered systems that may, in our opinion, be realized and addressed
in the context of physics of ultracold atomic or molecular gases. We
concentrate here mainly on fermionic systems.
This section is written on an elementary level and addressed to non-experts
in the physics of disordered systems.

\subsection{Anderson localization}

One of the most spectacular effects of disorder concerns single particles.
The spectrum of a Hamiltonian of a free particle in free space
or in a periodic lattice is
continuous and the corresponding eigenfunctions are extended (plane waves or
Bloch  functions). Introduction of disorder may drastically change this
situation. The basic knowledge about these phenomena comes
from the famous scaling theory formulated by the ``gang of four'' (\cite{gang},\cite{anderson}).

The scaling theory predicts that in 1D infinitesimally small disorder
leads to exponential localization of all eigenfunctions. The localization
length (defined as the width of the exponentially localized states) is a function of the ratio between the potential and the kinetic (tunneling) energies of the eigenstate and the disorder strength. For the case of discrete systems with constant tunneling
rates and local disorder distributed according to a Lorentzian distribution
(Lloyd's model, cf. \cite{haake}) the exact expression for the localization length is known.
In general, an exact relation between the density of states and the range of localization
in 1D has been provided by Thouless
\cite{thouless}. Hard core bosons with strong repulsion in 1D chains, described by $XY$ model
in a random  transverse field, can be mapped using the Wigner-Jordan transformation to 1D
non-interacting fermions in a random local potential,
which in turn maps the bosonic problem onto the problem of Anderson localization \cite{egger}.

In 2D, following the scaling theory, it is believed that localization occurs also for
arbitrarily small disorder, but its character interpolates smoothly between algebraic for weak disorder,
and exponential for strong disorder. There are, however, no rigorous arguments to support this belief,
and several controversies aroused about
this subject over the years. It would be evidently challenging to shed more light on this problem
using cold atoms in disordered lattices.

In 3D scaling theory predicts a critical value of disorder, above which every
eigenfunction exponentially localizes, and this fact has found strong evidence in numerical simulations.

In the area of AMO physics, effects of disorder have been studied in the context of
weak localization of light in random media \cite{kaiser}, which is believed to be a precursor of Anderson localization, and in the form of the so called
dynamical localization, that inhibits diffusion over the energy levels ladder in periodically
driven quantum chaotic systems, such as kicked rotor \cite{fishman}, microwaves driven hydrogen-like
atom (see \cite{haake} and references therein), or cold atoms kicked by optical lattices \cite{raizen}.

It is also worth mentioning at this point the existing large literature on unusual band structure and conductance
properties in systems with
incommensurate periodic potentials \cite{sokoloff}. The famous Harper's equation describing electron's
hopping in a $\cos(.)$
potential in 1D \cite{harper} may have, depending on
the strength of the potential, only localized, or only extended states due to the, so called, Aubry self-duality.
In more complicated cases without self-dual property, and/or in higher dimensions
coexistence of localized and extended states is very frequent.

\subsection{Localization in Fermi liquids}

The effects of disorder in electronic gases (i.e., Fermi gases with repulsive interactions) were in
the center of interest over many decades.
Originally, it was believed that weak disorder should not modify essentially the Fermi
liquid quasiparticle picture of Landau. Altshuler and Aronov \cite{altshuler}, and independently Fukuyama
\cite{fukuyama}, have shown, however,
that even weak disorder leads to surprisingly singular corrections to electronic
density of states near the Fermi surface, and to transport properties.

As we discussed above, for sufficient disorder in 3D all states are localized,
and the standard Fermi liquid theory is not valid. One can use then a Fermi-liquid like
theory using localized quasiparticle states. The system enters then an
insulating  {\it Fermi glass} state \cite{fermiglass}, termed often also as
Anderson insulator, in  which most of the interaction effects are included
in the properties of the Landau's quasiparticles.

In 1994 Shepelyansky has \cite{dima} stimulated further discussion about the role of interactions by
considering two interacting particles (TIP) in a random potential. He argued that two repulsing or attracting particles can propagate coherently on a distance much larger than the one-particle localization length.
Several groups have tried to study these effects of interplay between the disorder and (repulsive) interactions
in more detail in the regime when Fermi liquid becomes unstable as the Mott insulator state is approached by increasing the interactions. Numerical studies performed for spinless fermions with nearest neighbor (n.n.) interactions in a disordered mesoscopic ring;
for spin 1/2 electrons in a ring, described by the half-filled Hubbard-Anderson model;
for spinless fermions with Coulomb repulsion (reduced to n.n. repulsion) in 2D etc (\cite{metalglass},\cite{hofstetter}) show that as interactions become comparable with disorder, delocalization takes places.
In a 1D ring it leads to the appearance of persistent currents, in 2D the delocalized state exhibits also
an ordered flow of persistent currents, which is believed to constitute a novel quantum phase
corresponding to the metallic phase observed in experiments for instance with a gas of holes in
GaAs heterostructures for the similar range of parameters.

Another intensive subject of investigation concerns metal (Fermi liquid) - 
insulator transition driven by disorder in 3D. Theoretical description of this 
phenomenon goes back 
to the seminal works of Efros and Shklovskii \cite{efros}, and Mac Millan \cite{macmillan}.
 In this context, 
particularly impressive are the recent results of  
experiments on  disordered alloys, such as amorphous NbSi \cite{lee}, where 
the evidence for scaling and quantum critical behavior was found. Weakly 
doped semiconductors provide a good model of a disordered solid,  and their critical 
behavior at the metal-insulator transition has been intensively studied (cf. 
\cite{paal2}). Very interesting results concerning in  particular various forms 
of electronic glass: from 
Fermi glass, with negligible effects of Coulomb repulsion, to Coulomb glass \cite{coulomb}, 
dominated by the electronic correlations were obtained in the group of M. Dressel 
\cite{dressel}.

Although there exists experimental evidence for delocalization, enhanced persistent currents and novel metallic
phases at the frontier between the Fermi glass and Mott insulator, the further experimental models that
physics of cold atoms might provide are highly welcomed. Especially, since the cold atoms Hubbard toolbox
should allow to design with great fidelity
the models studied by theorists: spinless fermions extended Hubbard model in 1D, 2D and 3D, and
spin 1/2 Hubbard model in a disordered potential, or even more exotic systems such as Fermi systems with $SU(N)$ 'flavor' symmetry \cite{hofstetter2}. Perhaps a less ambitious, but still interesting
challenge is to use ultracold atomic gases to create both Fermi glass and a fermionic Mott insulator,
and investigate their  properties in detail.

\subsection{Localization in Bose systems}

At this point it is also worth mentioning the existing literature on the influence of repulsive
interactions on Anderson localization in bosonic systems.
In the weakly interacting case, one observes at low temperatures the phenomenon of Bose-Einstein condensation (BEC),
but to the eigenstates of the random potential (which are Anderson localized). Strong non-linear interactions tend, however,
to destroy the localization effects by introducing
screening of disorder by the non-linear mean field potential \cite{singh,rasmussen}. This happens as soon as the disorder 
localization length, $L_{dis}$, becomes larger than the healing length, $l_{heal}$. Such destruction of 
localization by weak nonlinearity
occurs also in the context of chaos, as discussed in 1993 by Shepelyansky \cite{dima93}. Several experiments, aiming at observation of
localization  with BEC's have been recently performed with elongated condensates in the presence of a speckle pattern and
1D optical lattices
\cite{inguscio,aspect,arlt}. In particular, transport suppression has been observed in the Orsay and Lens experiments, whereas,
as we have shown in the Hannover set-up \cite{arlt}, conditions for Anderson localization can be achieved
using additional
incommensurate superlattices. As the non-linearity (i.e., number of atoms) grows the condensate wave
function becomes a superposition
of exponentially localized modes of comparably low energies.
Overlapping of those modes signifies the onset of the screening regime. We believe that similar effects hold
in the strongly interacting limit in optical lattices, when they occur at the crossover from the Anderson glass ( in the weak interacting regime) to
Bose glass (in the strong interacting regime) behavior (see \cite{boseglass}, and also \cite{batrouni}).

\subsection{Localization in superconductors}

Obviously, the effects of disorder on superconductivity were studied practically from the very beginning of the theory of superconductors. Already in the late 50's
Anderson and Gorkov considered "dirty" superconductors \cite{dirty}. For a weak disorder, Bardeen-Cooper-Schrieffer (BCS) theory
is still valid,  but must be modified; the critical temperature is reduced by the localization
effects \cite{maekawa}.

The situation is more complex in the case of strong disorder. For example, in 2D superconductors
the superconducting state exists only for sufficiently small values of the disorder. This state is often
termed a superconducting vortex glass. Cooper pairs in this state condense and form a delocalized "Bose" condensate.
This condensate contains a large number of quantum vortices that are immobile and localized in the random
potential energy minima associated with disorder \cite{paalanen}. As disorder grows, the system enters
the insulating phase, which is a Bose glass of Copper pairs (for general theory see \cite{fisher}). Finally,
for even stronger values of the disorder the system enters
the insulating Fermi glass phase, when the Cooper pairs break down. Obviously,
this picture becomes even more complex at the BCS-BEC crossover.

Superconductor-insulator transition has been recently intensively studied in thin 
metal films on Ge or Sb substrates, that induce disorder on the atomic scale.
For not too thin films, transition 
to superconduting state occurs via Kosterlitz-Thouless-Berezinsky mechanism,
whereas for ultrathin films localization effects supress superconductivity \cite{liu,goldman}. 
In particular, scaling behavior and scaling exponent were studied  
in thin bismuth films \cite{bismuth}.

As before, the physics of cold gases might contribute here significantly to our
understanding of the influence of quenched disorder on the phenomenon of superconductivity.

\subsection{Localization and percolation}

Percolation is a classical phenomenon that is very closely related to localization \cite{cusack}.
Percolation provides a very general paradigm for a lot of
physical problems ranging from disordered electric devices \cite{grimmett},
forest fires and epidemics \cite{percbook,frisch}, to ferromagnetic ordering \cite{sachdev}.
In lattices, one considers site and bond percolation, and asks the following question: given
a probability of filling a
lattice site (filling a bond), and given a layer of the lattice of linear width $L$,
does exist a percolating  cluster of filled sites (bonds) that connects the walls of the layer?

Obviously, a percolating medium with a percolating cluster of empty sites
is an example of a  medium consisting of randomly  distributed scatterers. One has to expect that Anderson localization
will occur if quantum waves will propagate and scatter in such medium. An interplay between percolation and localization
has been a subject of intensive studies in the recent years. On one hand, when a classical flow is possible,
the quantum one might be suppressed due to the
destructive interferences and Anderson mechanism.
On the other hand, quantum  mechanics offers a possibility of tunneling through the
classically forbidden regions, and may thus allow for a classically forbidden flow. It turns out
that this latter mechanism is very weak, and one typically
observes three regimes of localization-delocalization behavior: classical localization below percolation limit,
quantum localization
above the percolation limit, and quantum delocalization sufficiently above the percolation
limit \cite{shapir,odagaki}.
Quantum percolation plays a role of mechanism responsible for quantum Hall effect \cite{ludwig}. Obviously,
interactions in the presence of quantum
percolation introduce additional complexity into the phenomenon.

Atomic Fermi-Bose mixtures and atomic gases in general offer an interesting possibility to study quantum percolation in a controlled way. One should stress that quenching atoms as random scatterers in a lattice
(below percolation threshold)
would be one of the methods itself to generate random local potentials.

\subsection{Random field Ising model}

Particularly interesting are those disordered systems, in which arbitrarily small disorder causes large qualitative effects,
with Anderson localization in 1D and, most presumably, in 2D being paradigm examples. Other examples are provided by
classical
systems that exhibit long range order at the lower critical dimension. In such systems, addition of an arbitrary small
local potential (magnetic field), that has a distribution assuming the same symmetry as the considered model,
destroys long range order.

The first example of such behavior has been shown by Imry and Ma \cite{imry}, using the domain wall argument;
it concerns random field Ising model in 2D,
for which magnetization vanishes  in
a random magnetic field in the Ising spin direction with symmetric distribution ($Z_2$ symmetry). This result has soon after
been proved rigorously \cite{proof}, and even generalized to $XY$, Heisenberg or Potts models (provided
that the corresponding "field" assumes the same symmetry as the model, i.e., $U(1)$, $SU(2)$ etc. \cite{wehr}).

One should note that most of the above discussed effects concern abstract spin models, and have no direct
experimental realizations in condensed matter systems. Cold atoms offer here a unique possibility of
both feasible realization of classical models, and of
studying quantum effects in those systems. Equally interesting in this context could be
spin models in which the random magnetic field breaks
the symmetry, such as for instance $XY$ model in 2D in the random field directed in, say, $X$ direction.
Such field breaks the $U(1)$ symmetry and changes the universality class of  the model to the Ising class.
Simultaneously, it prevents spontaneous magnetization in the $X$ direction. In effect, the system attains
the macroscopic magnetization in the $Y$ direction \cite{pelc}. We have recently studied these kind of  systems
and proved this result at $T=0$ rigorously. We expect in fact finite $T$ transition (as in Ising model),
but a detailed analysis of that case goes beyond the scope of the present paper \cite{wehr1}.

\subsection{Spin glasses - Parisi's theory and the ``droplet'' model}
Spin glasses are spin systems interacting via random couplings, that can be both ferro- or antiferromagnetic.
Such variations of the couplings lead typically to frustration. Spin glass models may thus exhibit many local minima of the free energy.
For this reason, spin glasses remain one of the challenges of the statistical physics and, in  particular, the question
of the nature of their ordering is
still open. According to Parisi's picture, spin glass phase consists of very many pure thermodynamic phases. The order
parameter of a spin glass becomes thus a function characterizing the probability of overlaps
between the distinct pure phases \cite{parisi}.  According to the, so called, "droplet" picture, developed by
Huse-Fisher and Bray-Moore \cite{stein,huse} there are (for Ising spin glasses) just two pure phases (up to $Z_2$ symmetry), and what frustration does is to change
very significantly the spectrum of excitations (domain walls, droplets) close to the equilibrium.
While the Parisi's picture (related also to the replica symmetry breaking) is most presumably valid for long range spin glass
models, such as Sherrington-Kirkpatrick model \cite{sherr}, the "droplet" model is formulated as
a scaling theory, and has a lot of numerical support for short range models, such as
Edwards-Anderson \cite{ea} model.

For more details of these two pictures relevant for our actual study see section V.

%%%%%%%%%%%%%%%%%%%%%%%%%%%%%%%%%%%%%%%%%%%%%%%%%%%%%%%%%%%%%
\section{Dynamics of composite fermions in the strong coupling regime}
\label{sec:3}
In this section we begin our detailed discussion of the low temperature physics of the Fermi-Bose mixtures.
In particular we consider a mixture of ultracold bosons (b) and spinless (or spin-polarized)
fermions (f), for example $^7$Li-$^6$Li or $^{87}$Rb-$^{40}$K, trapped in an optical lattice.
In the following, we will first consider the case of an homogeneous optical lattice, where all lattice
sites are equivalent, and we will review previous results focusing on formation of composite fermions and quantum
phase diagram \cite{lewen}. Second, we shall extend the analysis to the case of inhomogeneous optical lattices. We consider on-site inhomogeneities consisting in a harmonic confining potential and/or diagonal disorder.
In all cases considered below, the temperature is assumed to be low enough and the potential wells deep enough so that only quantum states in the fundamental Bloch band for bosons or fermions are populated. Note that, this requires that the filling factor for fermions $\rho_\textrm{f}$, is smaller than $1$, i.e., the total number of fermions, $N_\textrm{f}$, is smaller than the total number of lattice sites $N$.

In the tight-binding regime, it is convenient to project wavefunctions on the
Wannier basis of the fundamental Bloch band, corresponding to wavefunctions well
localized in each lattice site \cite{wannier,kittel2}. This leads to the Fermi-Bose Hubbard (FBH) Hamiltonian
\cite{Auerbach,sachdev,jaksch,other}~:
%+++++++++++++++++++++++++++++++++++%
\begin{eqnarray}
H_\textrm{FBH} & = &
-\sum_{\langle ij \rangle}\left[ J_\textrm{b} b^{\dagger}_i b_j+J_\textrm{f} f^{\dagger}_i f_j + h.c. \right] \nonumber \\
&& + \sum_i \left[\frac{V}{2} n_i(n_i-1) + U n_i m_i \right] \label{hamiltonian} \\
&& + \sum_i \left[-\mu^\textrm{b}_i n_i -\mu^\textrm{f}_i m_i \right] \nonumber
%+ \omega^\textrm{b}_i n_i+\omega^\textrm{f}_i m_i]
\end{eqnarray}
%+++++++++++++++++++++++++++++++++++%
where $b^{\dagger}_{i}$, $b_{i}$, $f^{\dagger}_{i}$ and $f_{i}$ are bosonic and fermionic creation- annihilation operators of a particle in the $i$-th localized Wannier state of the fundamental band and $n_i= b^{\dagger}_i b_i$, $m_i= f^{\dagger}_i f_i$ are the corresponding on-site number operators.
%; and $\mu^\textrm{b}_i$, $\mu^\textrm{f}_i$ are the bosonic and fermionic local chemical potentials; and $\omega^\textrm{b}_i$, $\omega^\textrm{f}_i$ are the bosonic and fermionic on-site potential energy.
The FBH model describes:
(i) nearest neighbor (n.n.) boson (fermion) hopping, with an associated negative energy,
$-J_\textrm{b}$ ($-J_\textrm{f}$);
(ii) on-site boson-boson interactions with an energy $V$, which is supposed to be positive (i.e., repulsive) in the reminder of the paper;
(iii) on-site boson-fermion interactions with an energy $U$, which is positive (negative) for repulsive (attractive) interactions;
(iv) on-site energy due to interactions with a possibly inhomogeneous potential, with energies $-\mu^\textrm{b}_i$ and $-\mu^\textrm{f}_i$; Eventually, $-\mu^\textrm{b}_i$ and $-\mu^\textrm{f}_i$ also contain the chemical potentials in grand canonical description.
% and (v) external harmonic confining potential with on-site energies $\omega^\textrm{b}_i$ and $\omega^\textrm{f}_i$.
For the shake of simplicity, we shall focus, in the following, on the case of equal hopping for fermions and bosons, $J_\textrm{b}=J_\textrm{f}=J$ and we shall assume strong coupling regime, i.e., $V,U \gg J$. Generalization to the case $J_\textrm{b} \not= J_\textrm{f}$ is just straightforward. 

%%%%%%%%%%%%%%%%%%%%%%%%%%%%%%%%%%%%%%%%%%%%%%%%%%%%%%%%%%%%%
\subsection{Quantum phases in homogeneous optical lattices}
\label{sec:homogeneous}

Before turning to inhomogeneous optical lattices, let us briefly review here the
results presented in \cite{lewen} for homogeneous lattices at zero temperature,
when all sites are translationally equivalent.
%-----------------------------------%
\begin{figure}
\includegraphics[width=1.1\linewidth]{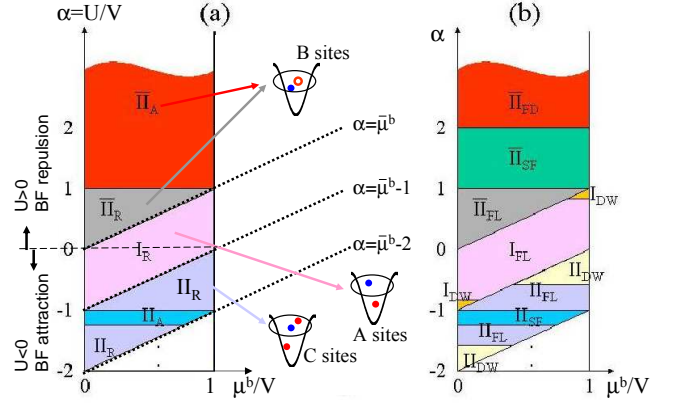}
\caption{(color online) Quantum phase diagrams of Fermi-Bose mixtures in an homogeneous optical 
lattice as functions of $\tilde{\mu}^\textrm{b}$ and $\alpha=U/V$, for $\rho_\textrm{f}=0.4$ and $J/V=0.02$. 
(a) Diagram of composites where the filled blue (red) dots symbolize fermions (bosons) and the red empty dots, bosonic holes. (b) Detailed quantum phase diagram of fermionic composites. 
}\label{fig:phase}
\end{figure}
%-----------------------------------%
In the limit of vanishing hopping ($J=0$) with finite repulsive boson-boson interaction $V$,
and in the absence of interactions between bosons and fermions ($U=0$), the bosons are in a Mott
insulator (MI) phase with exactly $\tilde{n}=\lceil\tilde{\mu}^\textrm{b}\rceil+1$ bosons per site,
where $\tilde{\mu}^\textrm{b}=\mu^\textrm{b}/V$ and $\lceil x \rceil$ denotes the
integer part of $x$. In contrast, the fermions can be in any set of Wannier states, since for vanishing tunneling,
the energy is independent of their configuration.
The situation changes when the interparticle interactions between bosons and fermions, $U$, are turned on. In the following, we define $\alpha=U/V$
and we consider the case of bosonic MI phase with $\tilde n$ boson per site. The presence of
a fermion in site $i$ may attract $-s > 0$ bosons or equivalently expel $s \leq \tilde n$ boson(s) depending on the sign of $U$.
The on-site energy gain in attracting $-s$ bosons or expelling $s$ bosons from site $i$ is
$\Delta E_i = \frac{V}{2}s(s-2 \tilde n + 1) - U s + \mu^\textrm{b} s$.
Minimizing $\Delta E_i$ it clearly appears energetically favorable to expel
$s=\left\lceil\alpha-\tilde{\mu}^\textrm{b}\right\rceil+\tilde{n}$ bosons.
Within the occupation number basis, excitations correspond of having $\tilde n -s \pm 1$ boson in a
site with a fermion, instead of $\tilde n -s$ and, therefore, the corresponding excitation energy is $\sim V$.
In the following, we assume that the temperature is smaller than $V$ so that the population of the
above mentioned excitations can be neglected. It follows that tunneling of a fermion is
necessarily accompanied by the tunneling of $-s$ bosons (if $s<0$) or
opposed-tunneling of $s$ bosons (if $s\geq 0$). The dynamics of our Fermi-Bose
mixture can thus be regarded as the one of composite fermions made of one fermion plus $-s$ bosons
(if $s<0$) or one fermion plus $s$ bosonic holes (if $s\geq 0$). The annihilation operators of the
composite fermions are \cite{lewen}:
%+++++++++++++++++++++++++++++++++++%
\begin{eqnarray}
F_i & = & \sqrt{\frac{(\tilde n -s)!}{\tilde n !}} \left( b_i^\dagger \right)^s f_i \quad\mbox{for $s$ bosonic holes}\label{eq:composites1} \\
F_i & = & \sqrt{\frac{\tilde n!}{(\tilde n -s)!}} \left( b_i \right)^{-s} f_i \quad\mbox{for $-s$ bosons} .\label{eq:composites2}
\end{eqnarray}
%+++++++++++++++++++++++++++++++++++%
These operators are fermionic in the sub-Hilbert space generated by $|n-ms,m\rangle$ with $m=0,1$ in
each lattice.
Note that within the picture of fermionic composites, the vacuum state corresponds to MI phase with
$\tilde n$ boson per site.
At this point, different composite fermions appear depending on the values of $\alpha$, $\tilde n$ and
$\tilde \mu^\textrm{b}$ as detailed in Fig.~\ref{fig:phase} \cite{lewen}.
The different composites are denoted by Roman numbers $I$, $II$, $III$, etc, which denote the
total number of particles that form the corresponding composite fermion. Additionally, a bar over a Roman
number indicates composite fermions formed by one bare fermion and bosonic holes, rather than bosons.
For the sake of simplicity, we shall consider only bosonic MI phases with $\tilde{n} = 1$ boson per site (i.e., $0<\tilde \mu^\textrm{b}\leq 1$) in the following parts of this paper \cite{foot1}.

If $\alpha - \tilde{\mu}^\textrm{b} > 0$, a fermion in site $i$ pushes the boson out of the site. We will call the sites with this
property $B$-sites. This notation will become particularly important in the presence of disorder (local
$\tilde{\mu}^\textrm{b}$). The composites, in this case,
correspond to one fermion plus one bosonic hole (this phase is called $\overline{II}$ in
Fig.~\ref{fig:phase}(a)). If $-1<\alpha - \tilde{\mu}^\textrm{b} < 0$, we have bare fermions as composites
(this corresponds to phase $I$). The sites with this property will be called $A$-sites. Finally,
if  $-2<\alpha - \tilde{\mu}^\textrm{b} < -1$,
the composites are made of one fermion plus one boson (phase $II$). The sites with the
latter property will be called $C$-sites.
Because all sites are equivalent for the fermions, the ground state is highly
$[N! / (N_\textrm{f})! (N - N_\textrm{f})!]$-degenerated, so the manifold of ground
states is strongly coupled by fermion or boson tunneling.
We assume now that the tunneling rate $J$ is small but finite. Using time-dependent
degenerate perturbation theory \cite{cohen}, we derive an effective Hamiltonian \cite{Auerbach}
for the fermionic composites:
%+++++++++++++++++++++++++++++++++++%
\begin{equation}
H_\textrm{eff}=-d_\textrm{eff}\sum_{\langle i,j \rangle}(F^{\dagger}_iF_j+ h.c.)+
K_\textrm{eff}\sum_{\langle i,j \rangle}M_iM_j - \overline{\mu}_\textrm{eff} \sum_i M_i
\label{Heff}
\end{equation}
%+++++++++++++++++++++++++++++++++++%
where
%$F^{\dagger}_i$ and $F_j$ are the corresponding composite fermionic annihilation and creation operators and
$M_i= F^{\dagger}_i F_i$ and $\overline{\mu}_\textrm{eff}$ is the chemical potential,
which value is fixed by the total number of composite fermions. The nearest neighbor hopping for
the composites is described by $-d_\textrm{eff}$ and the nearest neighbor composite-composite interactions
is given by $K_\textrm{eff}$, which may be repulsive ($>0$) or attractive ($<0$). This effective model is
equivalent to that of spinless interacting fermions. The interaction coefficient $K_\textrm{eff}$ originates
from $2$-nd order terms in perturbative theory and can be written in the general form~:
\begin{widetext}
%+++++++++++++++++++++++++++++++++++%
\begin{equation}
K_\textrm{eff} = \frac{-2J^2}{V} \left[
{(2\tilde{n}-s)(\tilde{n}+1)-s(\tilde{n}-s)}
-
\frac{(\tilde{n}-s)(\tilde{n}+1)}{1+s-\alpha}
-
\frac{(\tilde{n}-s+1)\tilde{n}}{1-s+\alpha}
-
\frac{1}{s\alpha}
\right] .
\label{Keff_hom}
\end{equation}
%+++++++++++++++++++++++++++++++++++%
\end{widetext}
This expression is valid in all the cases but when $s=0$, the last term ($1/s\alpha$) should not be taken into account. $d_\textrm{eff}$ originates from $(|s|+1)$-th order terms in perturbative theory and thus presents
different forms in different regions of the phase diagram of Fig.~\ref{fig:phase}. For instance in region I,
$d_\textrm{eff}=J$, in region $\overline{II}$ $d_\textrm{eff}=2 J^2 / U$ and
in region II, $d_\textrm{eff}=4 J^2/|U|$.

The physics of the system is determined by the ratio $K_\textrm{eff}/d_\textrm{eff}$ and the sign
of $K_\textrm{eff}$. In Fig.~\ref{fig:phase}(a), the subindex A/R denotes attractive
($K_\textrm{eff}>0$) / repulsive ($K_\textrm{eff}<0$) composites interactions. Fig.~\ref{fig:phase}(b)
shows the quantum phase diagram of composites for fermionic filling factor $\rho_\textrm{f}=0.4$ and tunneling $J/V=0.02$.
As an example, let us consider the case of repulsive interactions between bosons and fermions, $\alpha>0$.
Once the fermion-bosonic hole composites $\overline{II}$ have been created ($\alpha>\tilde \mu^\textrm{b}$),
the relation $K_\textrm{eff}/d_\textrm{eff}=-2(\alpha-1)$ applies. Consequently, if one increases the
interactions between bosons and fermions adiabatically, the system evolves through different quantum phases.
For $\tilde{\mu}^\textrm{b}<\alpha<1$, the interactions between composite fermions are repulsive and of
the same order of the tunneling; the system exhibits delocalized metallic phases. For $\alpha \simeq 1$,
the interactions between composite fermions vanish and the system show up the properties of an
ideal Fermi gas. Growing further the repulsive interactions between bosons and fermions,
the interactions between composite fermions become attractive. For  $1<\alpha<2$, one expects
the system to show superfluidity,  and for $\alpha>2$ fermionic insulator domains are predicted.

In the reminder of the paper,
we shall generalize these results to the case of inhomogeneous optical lattices. We shall assume diagonal
inhomogeneities, i.e., site-dependent local energies ($\mu_i^\textrm{b,f}$ depends on site $i$ but the
tunneling rate $J$ and interactions $U$ and $V$ do not). Diagonal inhomogeneities may account for
(i) overall trapping potential (usually harmonic), which is usually underlying in experiments
on ultracold atoms,  and (ii) disorder that may be introduced in different ways in
ultracold samples (see section~\ref{sec:experiments} for details).

%%%%%%%%%%%%%%%%%%%%%%%%%%%%%%%%%%%%%%%%%%%%%%%%%%%%%%%%%%%%%
\subsection{Composites in disordered lattices: effective Hamiltonian}
\label{sec:Heffective}

In this section we include on-site energy inhomogeneities in the optical lattice and we derive a generalized
effective Hamiltonian for the composite fermions.
Strictly speaking, in the presence of disorder the hopping terms should depend on the site under
consideration. Nevertheless, for weak enough disorder one can assume site independent tunneling for
both bosons or fermions \cite{boseglass}. In the following we will restrict ourselves
to the case where the hopping rates of bosons and fermions are equal and site-independent,
$J_b=J_f=J$ and to the strong coupling regime, $V,U>>J$, where the tunneling can be considered
as a perturbation, as in section~\ref{sec:homogeneous}.

For homogeneous lattices (see section~\ref{sec:homogeneous}), following the lines of
Refs.~\cite{svistunov,lewen}, we have used the method of degenerate second order
perturbation theory to derive the effective Hamiltonian (\ref{Heff}) by projecting the
wave function onto the multiply degenerated ground state of the system in the absence of tunneling.

In the inhomogeneous case, this approach cannot be applied since even
 for $J=0$ there exists a well defined single ground state determined by the values of the local chemical
potentials.
Nevertheless, in general, there will be a manifold of many states with similar energies. The differences of energy inside a manifold are of the order of the difference of chemical potential in different sites,
whose random
distribution is bounded, i.e., $0 \leq \tilde{\mu}^\textrm{b}_i$, $\tilde{\mu}^\textrm{f}_i \leq 1$. Moreover, the lower
energy manifold is
separated from the exited states by a gap given by the boson-boson interaction, $V$. Therefore, one can
apply a form of
quasidegenerate perturbation theory by projecting onto the manifold of near-ground states \cite{cohen}.

%This approach can be extended to inhomogeneous lattices where there are many states with very similar energies.
%We thus project the wavefunction on the manifold of ground states.
%In the presence of inhomogeneity this approach can not be applied since even for $J=0$ there exists a well defined single ground state determined by t%he values of the local chemical potentials, $\mu^\textrm{b}_i$ and $\mu^\textrm{f}_i$. Nevertheless, in general, there will be a manifold of many stat%es with similar energies.
%The differences of energy inside a manifold of ground states are of the order of the difference of on-site energies in different sites, whose random d%istribution has to be bounded to form the composites (see below).
%, i.e.  $0 \leq \tilde{\mu}^\textrm{b}_i$, $\tilde{\mu}^\textrm{f}_i \leq 1$.
%Moreover, the lower energy manifold is separated from the first exited states by a gap given by the boson-boson interaction, $V$. Therefore, one can a%pply the quasidegenerated perturbation theory projected onto the manifold of ground states \cite{cohen}.

As it is described in Refs. \cite{cohen} and briefly summarized in Appendix \ref{effhamil},
we construct an effective Hamiltonian that describes the slow, low-energy perturbation induced
within the manifold of unperturbed ground states by means of a unitary transformation applied to the total
Hamiltonian (\ref{hamiltonian}).
By denoting with P the projector on the manifold and Q=1-P its complement, the expression of the effective
Hamiltonian can be written as:
%+++++++++++++++++++++++++++++++++++%
\begin{eqnarray}
&&\langle out | H_\textrm{eff}|in\rangle =
\langle out |H_{0}|in\rangle + \langle out | PH_{int}P|in\rangle \label{hamil} \\
&&-\frac{1}{2}\langle out|
PH_{int}Q\left(\frac{1}{H_0-E_{in}}+\frac{1}{H_0-E_{out}}\right)
QH_{int}P|in\rangle.\nonumber
\end{eqnarray}
%+++++++++++++++++++++++++++++++++++%
As second order theory can only connect states that differ on one set of two adjacent sites,
the effective Hamiltonian
$H_\textrm{eff}$ can only contain nearest-neighbor hopping and interactions as well as
on-site energies $\overline{\mu}_i$ \cite{anna}:
%+++++++++++++++++++++++++++++++++++%
\begin{equation}
H_\textrm{eff}=\sum_{\langle i,j \rangle} \left[ -d_{i,j} F^{\dagger}_i F_j + h.c.\right] + \sum_{\langle i,j \rangle} K_{i,j} M_i M_j + \sum_i \overline{\mu}_i M_i
\label{Heffinhom}
\end{equation}
%+++++++++++++++++++++++++++++++++++%
where $M_i$, $F_i$ are defined as in (\ref{Heff}). The explicit calculation of the
coefficients $d_{i,j}$, $K_{i,j}$ and $\overline{\mu}_i$ depends on the concrete type of composites.
In the three following sections we address the cases of fermion-bosonic hole composites ($\overline{II}$),
bare fermion composites ($I$), and fermion-boson composites ($II$).

%%%%%%%%%%%%%%%%%%%%%%%%%%%%%%%%%%%%%%%%%%%%%%%%%%%%%%%%%%%%%
\subsection{Fermion-Bosonic hole composites} \label{FBhc}

In this section, we assume that all sites are $B$-sites, i.e.,  $\alpha-\tilde{\mu}_i^\textrm{b}>0$, so that
composite fermions $\overline{II}$ are created. This means that each site contains
either one boson or one fermion plus a bosonic hole. Thus, the manifold of low lying states comprises all possible configurations of $N_\textrm{f}$ fermions on an $N$-site lattice, with no
fermion occupied sites filled by bosons.

Within the manifold of ground states, a fermion jump from site $i$ to site $j$
can only occur if the boson that was initially in site $j$ jumps back to site $i$
%to fill the lack.
into the hole the fermion leaves behind.
 Therefore, the number operator for fermions and bosons are related with the number
operator of composites, i.e., $M_i=m_i=1-n_i$. Note that the composite model is expressed in terms of
the composite fermionic operators $F_i=b^{\dagger}_i f_i$ and thus $M_i=f^{\dagger}_i f_i  b_i b^{\dagger}_i$.
To determine the coefficients in (\ref{Heffinhom}), one looks
at two adjacent sites with indices $i$ and $j$
and uses a vector notation \ket{1_\textrm{b},1_\textrm{f}}\ which would correspond
to one boson on site $i$ and one fermion on site $j$.
In the composite fermion picture this would be denoted
as \cket{0,1}, i.e., one composite fermion on site $j$ and
no composite fermion on site $i$. With this notation
tunneling rates and nearest neighbor-interactions are calculated from Eq.~(\ref{hamil}) as:
\begin{widetext}
\begin{eqnarray}\bra{1_\textrm{f},1_\textrm{b}}\Heff\ket{1_\textrm{b},1_\textrm{f}}&=&
-\frac{1}{2}\JsqoverV \lrb \frac{1}{\alpha-\Deltaij^\textrm{b}}
+\frac{1}{\alpha+\Deltaij^{\textrm{b}}}+\frac{1}{\alpha-\Deltaij^{\textrm{f}}}
+\frac{1}{\alpha+\Deltaij^{\textrm{f}}}\rrb \nonumber\\
&\equiv&\cbra{1,0}\aiaj{F}\cket{0,1} \label{brak1}
\end{eqnarray}
\begin{eqnarray}\bra{1_\textrm{b},1_\textrm{b}}\Heff\ket{1_\textrm{b},1_\textrm{b}}&=&
-\frac{1}{2}\JsqoverV\lrb\frac{2}{1-\Deltaij^\textrm{b}}
+\frac{2}{1+\Deltaij^\textrm{b}}+\frac{2}{1-\Deltaij^\textrm{b}}
+\frac{2}{1+\Deltaij^\textrm{b}}\rrb \nonumber\\
&\equiv&\cbra{0,0}(1-\Mi)(1-\Mj)\cket{0,0}\label{brak2}
\end{eqnarray}
\begin{eqnarray}\bra{1_\textrm{f},1_\textrm{b}}\Heff\ket{1_\textrm{f},1_\textrm{b}}&=&
-\frac{1}{2}\JsqoverV\lrb \frac{2}{\alph-\Deltaij^\textrm{b}}
+\frac{2}{\alph+\Deltaij^\textrm{f}}\rrb \nonumber\\
&\equiv&\cbra{1,0}\Mi(1-\Mj)\cket{1,0}\label{brak3}
\end{eqnarray}
\begin{eqnarray}\bra{1_\textrm{b},1_\textrm{f}}\Heff\ket{1_\textrm{b},1_\textrm{f}}&=&
-\frac{1}{2}\JsqoverV\lrb \frac{1}{\alph+\Deltaij^\textrm{b}}
+\frac{1}{\alph-\Deltaij^\textrm{f}}\rrb \nonumber\\
&\equiv&\cbra{1,0}\Mi(1-\Mj)\cket{1,0}\label{brak4}
\end{eqnarray}
\end{widetext}
Summing these terms up yields the coefficients for 
(\ref{Heffinhom}):
\begin{widetext}
%+++++++++++++++++++++++++++++++++++%
\begin{eqnarray}
d_{ij}&=&\frac{J^2}{V}\left(\frac{\alpha}{\alpha^2-(\Delta^\textrm{b}_{ij})^2}
+\frac{\alpha}{\alpha^2-(\Delta^\textrm{f}_{ij})^2}\right)\label{dij2} \\
K_{ij}&=&-\frac{J^2}{V}\Bigg(\frac{4}{1-(\Delta^\textrm{b}_{ij})^2}
-\frac{2\alpha}{\alpha^2-(\Delta^\textrm{f}_{ij})^2}
-\frac{2\alpha}{\alpha^2-(\Delta^\textrm{b}_{ij})^2}\Bigg) \label{kij2} \\
\overline{\mu}_i &=& \mu_i^{\textrm{b}}-\mu_i^{\textrm{f}}
+ \frac{J^2}{V} \left[ \sum_{\langle i,j \rangle}
{\frac{4}{1-(\Delta_{ij}^{\textrm{b}})^2}}-
{\frac{1}{\alpha-\Delta_{ij}^{\textrm{b}}}}-
{\frac{1}{\alpha+\Delta_{ij}^{\textrm{f}}}}
\right] \label{muij2}
\end{eqnarray}
%+++++++++++++++++++++++++++++++++++%
\end{widetext}
with $\Delta_{ij}^\textrm{f,b}=\tilde{\mu}_i^\textrm{f,b}-\tilde{\mu}_j^\textrm{f,b}$.
Here, $\langle i,j \rangle$ represents all neighbor sites of $i$. We shall now consider separately two limiting cases: (i) $\mu_i^\textrm{f}=0$  and $\mu_i^\textrm{b}=\mu_iV$,
and (ii) $\mu_i^\textrm{f}=\mu_i^\textrm{b}=\mu_iV$ .

\subsubsection{Case where $\mu_i^\textrm{f}=0$}\label{iiiB1}
\begin{figure}
\includegraphics[width=1.0\linewidth]{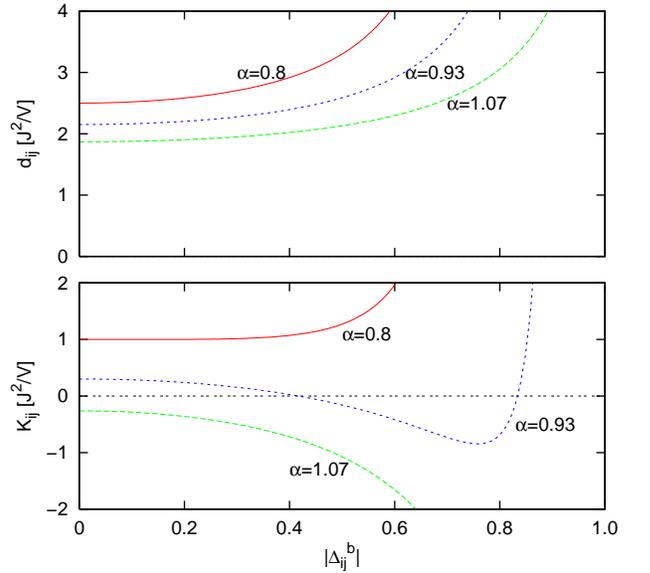}
\caption{Tunneling, $d_{ij}$, and nearest neighbor couplings, $K_{ij}$, of type $\overline{II}$ composites as
a function of disorder $\Delta_{ij}^\textrm{b}$, and for different boson-fermion interactions, $\alpha$, in the case
$\mu_i^\textrm{f}=0$.}\label{fig:sitesB}
\end{figure}

In the first case, we assume that the on-site energy for fermions vanishes.
We assume also that all sites are $B$-sites, i.e., $\alpha-\tilde{\mu}_i^\textrm{b}>0$ everywhere.
In this case, the hopping amplitudes $d_{ij}$ are always positive, although may vary quite significantly with disorder, especially when $\Delta_{ij}^\textrm{b}\simeq\alpha$.
As shown in Fig.~\ref{fig:sitesB}, for $\alpha>1$, $K_{ij}\le 0$ and we deal with attractive (although random) interactions. For $\alpha<1$, $K_{ij}\ge 0$ and the interactions between composites are repulsive. For $\alpha<1$, but close to 1, $K_{ij}$ might take positive or negative values for $\Delta_{ij}^\textrm{b}$ small or
$\Delta_{ij}^\textrm{b}\simeq \alpha$. In this case, the {\it qualitative character of interactions}
may be {\it controlled by inhomogeneity} \cite{anna}. At low temperatures the physics of the system will depend on the relation between  $\tilde{\mu}_i^\textrm{b}$'s and $\alpha$.
%-----------------------------------%
%\begin{figure}[t!]
%\mbox{\epsfxsize 3.0in \epsfbox{./dK2bar.eps}}
%\caption{Tunneling $d_{ij}$ and nearest neighbor couplings $K_{ij}$ of type $\overline{II}$ %composites as
%a function of disorder $\Delta_{ij}^\textrm{b}$, and for different boson-fermion interactions in %the case
%$\mu_i^\textrm{f}=0$.}
%\label{fig:sitesB}
%\end{figure}

\begin{figure}
\includegraphics[width=1.0\linewidth]{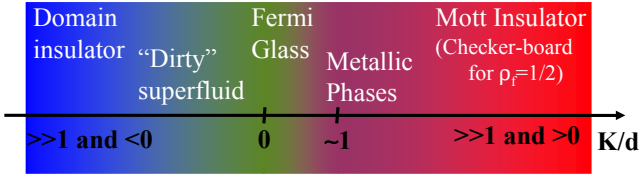}
\caption{(color online) Schematic phase diagram of type $\overline{II}$ fermionic composites
for low disorder ($\Delta_{ij}^\textrm{b} \ll 1, \alpha$) and vanishing fermionic on-site
energy ($\mu_i^\textrm{f}=0$) as a function of the ratio between n.n. interactions and tunneling for the composites.}\label{fig:sitesBbis}
\end{figure}

%-----------------------------------%
\subparagraph*{Small disorder limit -}
For small disorder, we may neglect the contributions of $\Delta_{ij}^\textrm{b}$ to $d_{ij}\simeq d$ and
$K_{ij}\simeq K$, and keep only the leading disorder contribution in $\overline\mu_i$, i.e., the
first term in Eq.~(\ref{muij2}). Note, that the latter contribution is relevant in 1D and 2D
leading to Anderson localization of single particles \cite{gang}.
When $K/d\ll 1$ the system will then be in the Fermi glass phase,
i.e., Anderson localized (and many-body corrected) single particle
states will be occupied according to the Fermi-Dirac
rules \cite{fermiglass}. For repulsive interactions and $K/d\gg 1$,
the ground state will be a Mott insulator
and the composite fermions will be pinned for large filling factors. In particular, for
filling factor $\rho_\textrm{f}=1/2$, one expects the ground state to be in the form of a checker-board.
For intermediate values of $K/d$, with $K>0$, delocalized metallic phases with enhanced
persistent currents are possible \cite{metalglass}.
Similarly, for attractive interactions ($K<0$) and  $|K|/d< 1$ one expects competition between
pairing of fermions
and disorder, i.e., a "dirty" superfluid phase while for $|K|/d\gg 1$, the fermions will form a domain insulator.
Fig.~\ref{fig:sitesBbis} shows a schematic representation of expected disordered phases of
the type $\overline{II}$ fermionic composites
for small disorder, and
vanishing fermionic on-site chemical potential.
%-----------------------------------%
%-----------------------------------%
\subparagraph*{Spinglass limit -}
Another interesting limit corresponds to the case $\Delta_{ij}^\textrm{b}\simeq\alpha\simeq 1$. Such a situation
can be achieved by combining a superlattice potential with a spatial period twice as large as the
one of the lattice (which alone induces $|\Delta_{ij}^\textrm{b}|=1$) and a random potential to induce site-to-site
fluctuations.
The tunneling becomes then non-resonant and can be neglected in Eq.~(\ref{Heffinhom}),
while the couplings $K_{ij}$ fluctuate strongly as shown in Fig.~\ref{fig:sitesB}.
We end up then with the (fermionic) Ising spin glass model \cite{anna} described by the Edwards-Anderson
model with $s_i=2M_i-1=\pm 1$. This case is studied in more detail in section~\ref{sec:spinglass}.
%-----------------------------------%
%\begin{figure}[t!]
%\mbox{\epsfxsize 3.0in \epsfbox{./dK2mufermion.eps}}
%\caption{Tunneling $d_{ij}$ and nearest neighbor couplings $K_{ij}$ of type $\overline{II}$ composites as
%a function of disorder $\Delta_{ij}^\textrm{b}$, and for different boson-fermion interactions in the case
%$\mu_i^\textrm{f}=\mu_i^\textrm{b}=\mu_iV$.}
%\label{fig:sitesBmuf}
%\end{figure}

\begin{figure}
\includegraphics[width=1.0\linewidth]{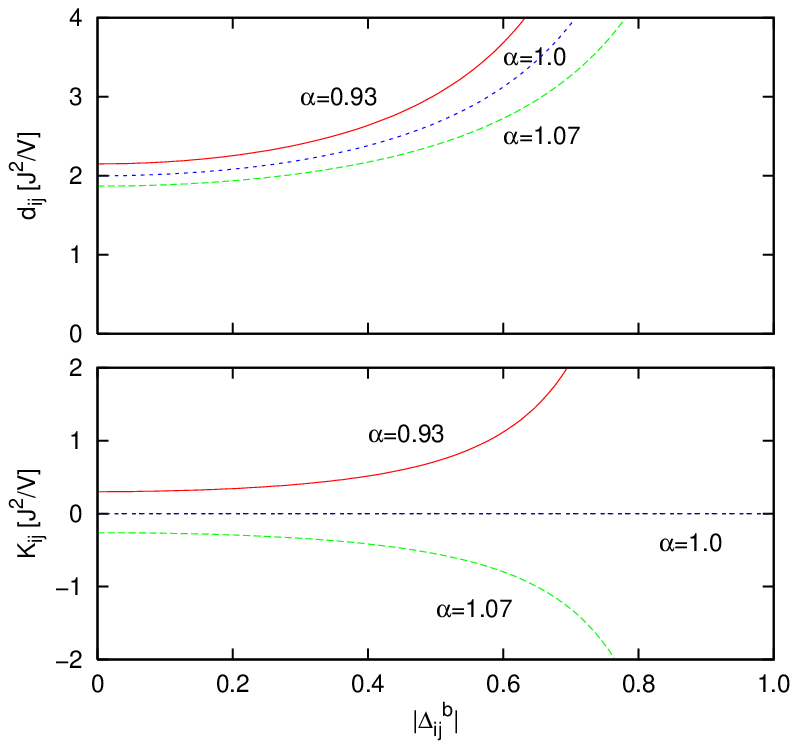}
\caption{Tunneling, $d_{ij}$, and nearest neighbor couplings, $K_{ij}$, of type $\overline{II}$ composites as
a function of disorder $\Delta_{ij}^\textrm{b}$, and for different boson-fermion interactions, $\alpha$, in the case
$\mu_i^\textrm{f}=\mu_i^\textrm{b}=\mu_iV$.}\label{fig:sitesBmuf}
\end{figure}
%-----------------------------------%
\subsubsection{Case where $\mu_i^\textrm{f}=\mu_i^\textrm{b}$}
%{\bf TO BE CONTROLLED / DEVELOPED:}
Let us now consider that the chemical potential is equal for bosons and fermions
at each lattice site, $\mu_i^\textrm{f}=\mu_i^\textrm{b}=\mu_iV$. All sites are still assumed to be $B$-sites.

%In this limit, the B sites are not the sites energetically favorable but A sites.
The effective interactions are for $\alpha>1$ always negative, and therefore the composites
experience random attractive interactions (as in the previous case),
while for $\alpha<1$, $K_{ij}>0$,  and therefore we deal with random repulsive interactions.
For $\alpha=1$, the interactions between composites vanish for all the values of the amplitude of the disorder.

In this case the sign of the interactions between composites is governed by the interactions
between bosons and fermions {\it alone}. Note that this is not possible, when one considers only disorder for the bosons. Fig.~\ref{fig:sitesBmuf} shows the tunneling and the nearest neighbor couplings for different values of $\alpha$. We expect here the appearance of similar phases, as in the previously discussed case.

%%%%%%%%%%%%%%%%%%%%%%%%%%%%%%%%%%%%%%%%%%%%%%%%%%%%%%%%%%%%%
\subsection{Bare Fermion composites}
In this section we now assume that all sites are $A$-sites and correspond to type $I$ fermionic composites, i.e.,
$-1<\alpha - \tilde{\mu}^\textrm{b} _i  < 0$. This means that composite fermions reduce to bare
fermions ($F_i=f_i$) flowing on the top of a MI phase with $\tilde n = 1$ boson per site. Each site
contains now one boson plus eventually one fermion. From application of perturbation theory as described in section~\ref{sec:Heffective}
[see Eq.~(\ref{hamil})], one finds that  the coefficients of the effective
Hamiltonian (\ref{Heffinhom}) are:
\begin{widetext}
%+++++++++++++++++++++++++++++++++++%
\begin{eqnarray}
d_{ij}&=&J ,\label{dij1}\\
K_{ij}&=&-\frac{J^2}{V}\left(
  \frac{8}{1-(\Delta^\textrm{b}_{ij})^2}
- \frac{4(1+\alpha)}{(1+\alpha)^2-(\Delta^\textrm{b}_{ij})^2}
- \frac{4(1-\alpha)}{(1-\alpha)^2-(\Delta^\textrm{b}_{ij})^2} \right) ,\label{kij1} \\
\overline{\mu}_i&=&-\mu_i^\textrm{f}
+ \frac{J^2}{V} \sum_{\langle i,j \rangle}
\left[
{\frac{4}{1-(\Delta_{ij}^{\textrm{b}})^2}}-
{\frac{1}{\Delta_{ij}^{\textrm{f}}}}-
{\frac{4}{1-(\alpha-\Delta_{ij}^{\textrm{b}})^2}}
\right] ~. \label{muij1}
\end{eqnarray}
%+++++++++++++++++++++++++++++++++++%
\end{widetext}
We observe that the inhomogeneities for fermions (site-dependent $\mu_i^\textrm{f}$)
do not neither perturb the effective tunneling, nor the effective interaction parameter,  while
$\overline\mu_i\simeq -\mu_i^\textrm{f}$ up to corrections of the order of $\textrm{O}(J^2/V)$ for
type I composite (bare) fermions. In this case, composite tunneling $d_{ij}$ originates from the first order term, while the nearest
neighbor interaction originates from second order perturbation. It should be noted that in the case of type $I$ composites, the hopping $d_{ij}$ and
interaction $K_{ij}$ parameters in Eq.~(\ref{Heffinhom}) do not depend on the sign of the
fermion-boson interaction $\alpha$. 
%-----------------------------------%
%\begin{figure}[t!]
%\mbox{\epsfxsize 3.0in \epsfbox{./dK1.eps}}
%\caption{Ratio of nearest neighbor coupling and tunneling $K_{ij}/d_{ij}$ as
%a function of disorder $\Delta_{ij}^\textrm{b}$ for $\alpha=0.7$ and $J/V=0.01$ in the case of type
%$I$ composite fermions.}
%\label{fig:sitesA}
%\end{figure}

\begin{figure}
\includegraphics[width=1.0\linewidth]{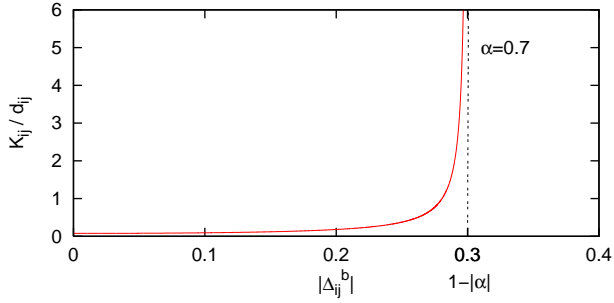}
\caption{Ratio of nearest neighbor coupling and tunneling $K_{ij}/d_{ij}$ as
a function of disorder $\Delta_{ij}^\textrm{b}$ for $\alpha=0.7$ and $J/V=0.01$ in the case of type
$I$ composite fermions.}\label{fig:sitesA}
\end{figure}

\begin{figure}
\includegraphics[width=0.4\linewidth]{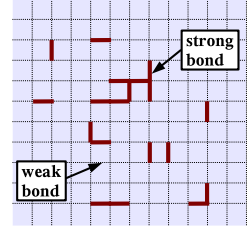}
\caption{Schematic representation of connecting bonds in type $I$ composite systems. The bonds are
separated in two types: (i) the weak bonds in which two composites do not interact and (ii) the strong
bonds where only one composite can stay. The short lines represent the bonds and the crossing points of
the bonds are the lattice sites.}\label{fig:percol}
\end{figure}

%-----------------------------------%
The couplings $K_{ij}$ are always positive, and for $\alpha\simeq 0$, $K_{ij}\simeq \textrm{O}(\alpha^2)$,
and both the
repulsive interactions, and disorder are very weak, leading to an almost ideal Fermi liquid behavior at
low temperature.
For finite $\alpha$, and $\Delta_{ij}^\textrm{b}\simeq 1-|\alpha|$, however, the fluctuations of $K_{ij}$ might be
quite large as shown in Fig.~\ref{fig:sitesA}.
Note, that for $|\alpha|\simeq 1$, this will occur even for small disorder.
It is interesting to note that the dynamics of type $I$ composites in our system resembles
{\it quantum bond percolation}.  As suggested from Fig.~\ref{fig:sitesA}, one can assume in
a somehow simplified view that the interaction parameter $K_{ij}$ takes either very large, or
zero values. The lattice decomposes into two sub-lattices (see Fig.~\ref{fig:percol}): a "weak" bond
sub-lattice (corresponding to $K_{ij} \ll d_{ij}$) in which fermions flow as
in an almost ideal Fermi liquid, and a "strong" sub-lattice (corresponding to $K_{ij} \gg d_{ij}$),
where only one fermion per bond is allowed ($M_i M_j =0$ for all nearest neighbors in the "strong" cluster).
Therefore, we see that the physics of bond percolation \cite{percbook,grimmett} will play a role.
For $p>p_\textrm{c}$, where $p$ is the density of weak bonds and $p_\textrm{c}\simeq 0.50$
(in two-dimensional square lattices) and $p_\textrm{c}\simeq 0.25$ (in three-dimensional cubic lattices),
the weak bond sub-lattice will be percolating, i.e., there exists a large cluster of weak bonds which spans
the lattice from one side to the other. The question arises as to determine the quantum bond percolation
threshold $p_\textrm{Q}$, i.e., for which minimal value of $p$, the eigenstates of the quantum gas will
be delocalized over the extension of the system. Although it is clear that $p_\textrm{Q} > p_\textrm{c}$,
it is still an open question to determine the exact value for the quantum percolation threshold
$p_\textrm{Q}$ \cite{shapir,chang,root,soukoulis91,soukoulis92}. Therefore, experimental realization
of our system may be of considerable interest for addressing this general question.
%-----------------------------------%
%\begin{figure}[t!]
%\mbox{\epsfxsize 2.00in \epsfbox{./percol.eps}}
%\caption{Schematic representation of connecting bonds in type $I$ composite systems. The bonds are
%separated in two types: (i) the weak bonds in which two composites do not interact and (ii) the strong
%bonds where only one composite can stay. The short lines represent the bonds and the crossing points of
%the bonds are the lattice sites.}
%\label{fig:percol}
%\end{figure}

%-----------------------------------%
%%%%%%%%%%%%%%%%%%%%%%%%%%%%%%%%%%%%%%%%%%%%%%%%%%%%%%%%%%%%%
\subsection{Fermion-Boson composites}
We finally consider in this section the case, when all sites are $C$-sites, so that type II composites
corresponding to $-2<\alpha - \tilde{\mu}_i^\textrm{b} < -1$ are formed.
The composites are made of one fermion and one boson. This means that each lattice site is populated
by either one boson or one fermion plus two bosons. Tunneling as well as nearest neighbor interaction
of composites arise from 2nd order terms in perturbative theory [see section~\ref{sec:Heffective} and
Eq.~(\ref{hamil}) for details]. Along the lines of section~\ref{sec:Heffective}, we find the
following expressions:

\begin{widetext}
%+++++++++++++++++++++++++++++++++++%
\begin{eqnarray}
d_{ij}&=&\frac{J^2}{V}\left(\frac{2\vert\alpha\vert}{\vert\alpha\vert^2-(\Delta^\textrm{b}_{ij})^2}
+\frac{2\vert\alpha\vert}{\vert\alpha\vert^2-(\Delta^\textrm{f}_{ij})^2}\right) ,\label{dij3}\\
K_{ij}&=&\frac{-J^2}{V}\left(\frac{16}{1-(\Delta_{ij}^\textrm{b})^2}-
\frac{2\vert \alpha \vert}{\vert \alpha \vert^2-{( \Delta_{ij}^\textrm{f} )}^2}
-\frac{8\vert \alpha \vert}{\vert \alpha \vert^2-{( \Delta_{ij}^\textrm{b} )}^2}
-\frac{6(2-\vert \alpha \vert)}{(2-\vert \alpha \vert)^2-{( \Delta_{ij}^\textrm{b})}^2}\right),\label{kij3} \\
\overline{\mu}_i &=& -\mu_i^\textrm{f}-\mu_i^\textrm{b}
+ \frac{J^2}{V} \sum_{\langle i,j \rangle}
\left[
{\frac{4}{1-(\Delta_{ij}^{\textrm{b}})^2}}-
{\frac{4}{|\alpha|+\Delta_{ij}^{\textrm{b}}}}
-{\frac{3}{2-|\alpha|-\Delta_{ij}^{\textrm{b}}}}-
{\frac{1}{|\alpha|+\Delta_{ij}^\textrm{f}}}
\right] ~. \label{muij3}
\end{eqnarray}
%+++++++++++++++++++++++++++++++++++%
\end{widetext}

Different scenarios also arise in this case. In the following, we shall
consider the case  $\mu_i^\textrm{f}=0$ and $\mu_i^\textrm{b}=\mu_iV$. The other extreme case,
$\mu_i^\textrm{f}=\mu_i^\textrm{b}=\mu_iV$, leads to qualitatively similar conclusions.

\subsubsection{Case where $\mu_i^\textrm{f}=0$}
\begin{figure}
\includegraphics[width=1.0\linewidth]{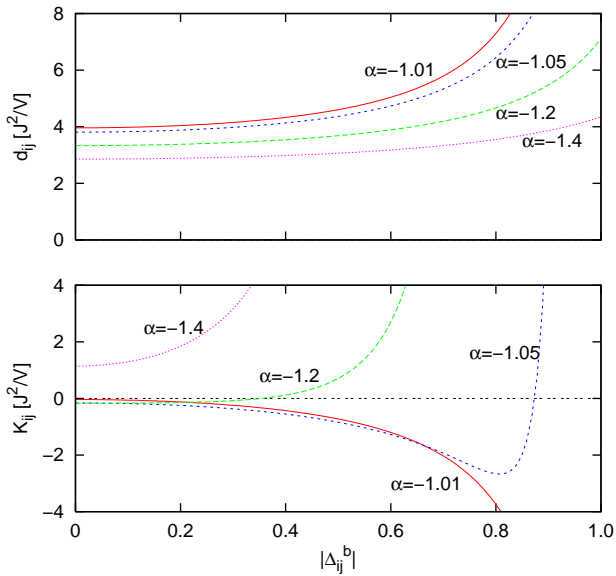}
\caption{Tunneling, $d_{ij}$, and nearest neighbor couplings, $K_{ij}$, of type ${II}$ composites as
a function of disorder $\Delta_{ij}^\textrm{b}$ and for different boson-fermion interactions, $\alpha$, in the case $\mu_i^\textrm{f}=0$.}\label{fig:sitesC}
\end{figure}

We assume here that the on-site energy for fermions is $\mu_i^\textrm{f}=0$.
As for Fig.~\ref{fig:sitesB}, we plot the effective tunneling and interaction parameter versus inhomogeneity parameter $\Delta_{ij}^\textrm{b}$ in Fig.~\ref{fig:sitesC}. The general behavior of
$d_{ij}$ and $K_{ij}$ is qualitatively the same as in the case of type $\overline{II}$ composites.
For type $II$ composites, and for small disorder, we find $K/d=1-4|\alpha|+3/(2-|\alpha|)$ with $1<|\alpha|<2$. The inhomogeneity is now given, for type $II$
composites, by $\overline{\mu}_i=-\mu_i^\textrm{b}$. The regimes, where $K \ll d$ corresponding to
an almost ideal Fermi gas (in the absence of disorder), or to a Fermi glass (in the presence of disorder),
can be reached in the region $\alpha \simeq -5/4$. The opposite regimes of strong effective interactions,
where $K \gg d$ appears for $\alpha \gtrsim -2$, and corresponds to repulsive interactions $K>0$.
In this region, the fermionic checkerboard phase if the filling factor is $1/2$ (for vanishing disorder)
and the repulsive Fermi glass phase (in the presence of disorder) are expected. Here, no strong attractive
interactions regime occurs since $K/d$ reaches a minimum of $\simeq -0.07$ for $\alpha=\sqrt{3}/2-2$. Therefore,
in contrast to type $\overline{II}$ composites, for type ${II}$ composites: (i) due to weakness of attractive
interactions, the domain insulator phase does not
appear, and even the "dirty" superfluid phase may be washed out; and (ii) arbitrary strong repulsive interactions
can be used to generate a Mott insulator, which might be difficult  for $\overline{II}$ composites, where
$K/d$ is limited to $2$, suggesting that Fermi liquid, Fermi glass behavior will prevail.
%-----------------------------------%
%\begin{figure}[t!]
%\mbox{\epsfxsize 3.0in \epsfbox{./dK2.eps}}
%\caption{Tunneling $d_{ij}$ and nearest neighbor couplings $K_{ij}$ of type ${II}$ composites as
%a function of disorder $\Delta_{ij}^\textrm{b}$ and for different boson-fermion interactions.}
%\label{fig:sitesC}
%\end{figure}

%-----------------------------------%
As shown in Fig.~\ref{fig:sitesC}, the spinglass limit can also be reached
for example for $\alpha \simeq -1.05$ and $|\Delta_{ij}^\textrm{f}| \simeq 0.9$. In this regime,
the tunneling is non-resonant due to strong disorder and the nearest neighbor interaction
fluctuates strongly from negative to positive values. See section~\ref{sec:spinglass} for
further study of the spinglass limit.
% ALL SITES D: RESULTS PAPER ANNA SITES A. Do not affect the chemical potential for the fermions
%+++++++++++++++++++++++++++++++++++%
%\begin{eqnarray}
%d_{ij}&=&J\label{dij4}\\
%K_{ij}&=&-\frac{-J^2}{V}(\frac{8}{1-(\Delta_{ij})^2}-
%\frac{4(1+\vert \alpha\vert)}{(1+\vert \alpha\vert)^2
%-{(\Delta_{ij})}^2}\\&-&
%\frac{4(1-\vert \alpha\vert)}{(1-\vert \alpha\vert)^2-(\Delta_{ij})^2}).\label{kij4}
%\end{eqnarray}
%+++++++++++++++++++++++++++++++++++%
%%%%%%%%%%%%%%%%%%%%%%%%%%%%%%%%%%%%%%%%%%%%%%%%%%%%%%%%%%%%%
\subsection{Optical lattices with different types of sites}
\subsubsection{Sites A and B}

Obviously, the situation becomes much more complex when we deal with different types of sites in the lattice.
Again there are infinitely many possibilities, and the simplest ones are, for instance: i) coexistence of $A$- and $B$-sites,
or ii) $A$- and $C$- sites, or iii) $A$-, $B$-, and $C$-sites, etc.
In the following we shall consider only the case i) with $\mu_i^\textrm{f}=0$  and $\mu_i^\textrm{b}=\mu_iV$, since the other cases lead to qualitatively similar effects.

Let us assume that the numbers of $A$-, and $B$- sites are macroscopic, i.e., of the order of $N$. More precisely, we will consider that
$N_A$ (number of $A$-sites) and $N_B$ (number of $B$-sites) of order N/2. In this case the physics of site percolation \cite{percbook} will play a role.
If $N_\textrm{f}\le N_B$ the composite fermions will move
within a cluster of $B$-sites. When $N_B$ will be above
the classical percolation threshold, this cluster will be percolating.
The expressions Eq. (\ref{dij2}) and Eq. (\ref{kij2}) will still be valid,
 except that they will connect only the $B$-sites.
The physics of the system will be similar as in the case of type $I$ composites),
but it will occur now
on the percolating cluster. For small disorder,
and  $K/d\ll 1$ the system will be in a Fermi glass phase
in which the interplay between the   Anderson localization
of single particles due to fluctuations of $\mu_i^\textrm{b}$ and quantum percolation
effects, that is randomness of the $B$-sites cluster, will occur.
For repulsive interactions and
$K/d\gg 1$, the ground state will be
a Mott insulator on the cluster
and the composite fermions  will be pinned (in particular
for half-filling of the cluster). It is an open question whether
the delocalized  metallic phases with enhanced
persistent current of the kind discussed in Ref. \cite{metalglass}
might exist in this case. Similarly, it is an open question whether
for attractive
interactions ($K<0$) and  $|K|/d< 1$
pairing of (perhaps localized) fermions
will take place. In the case of $|K|/d\gg 1$, we expect that  the fermions
will form a domain insulator on the cluster.

In the "spin-glass" limit $\Delta_{ij}^\textrm{b}\simeq\alpha\simeq 1$,
we will deal with the Edward-Anderson spin glass on the cluster.
Such systems are of interest in condensed matter physics
(cf. \cite{spiperc}), and again questions connected to the
nature of spin glass
ordering may be studied in this case.

When $N_\textrm{f}>N_B$, all $B$-sites will be filled, and the physics will occur on
the cluster of $A$-sites.
For $\alpha\simeq 0$, we will deal
with a gas with very weak repulsive interactions, and no significant
disorder on the random cluster; this is an ideal test to study
quantum percolation at low T.
For finite $\alpha$, and $\Delta_{ij}^\textrm{b}\simeq
1-\alpha$, the interplay between the fluctuating repulsive $K_{ij}$'s
and quantum percolation might be studied.

%%%%%%%%%%%%%%%%%%%%%%%%%%%%%%%%%%%%%%%%%%%%%%%%%%%%%%%%%%%%%
\section{Numerical analysis using Gutzwiller ansatz}
\label{sec:4}
\subsection{Numerical method}
In this section, we present numerical results that give evidences of (i) formation of composite
particles in Fermi-Bose mixtures in optical lattices and (ii) existence of different quantum phases
for various sets of composite tunneling and interaction parameters and inhomogeneities. We mainly focus
on type $\overline{II}$ composites.
Mean field theory provides appropriate although not exact properties of Hubbard models \cite{sachdev}.
In the following, we consider a variational mean field approach provided by the Gutzwiller ansatz
(GA) \cite{fisher,statGA}.
In particular, the GA ansatz has been successfully employed for bosonic systems to study
the superfluid to Mott Insulator transition \cite{fisher,jaksch} in non-disordered lattices,
and the Anderson and Bose glass transitions in the presence of disorder \cite{fisher,boseglass}.

Briefly, the Gutzwiller approach neglects site-to-site quantum coherences so that
the many-body ground state is written as a product of $N$ states, each one being localized in a
different lattice site. Each localized state is a superposition of different Fock states $|n,m\rangle_i$
with exactly $n$ bosons and $m$ fermions on the $i$-th lattice site~:

%+++++++++++++++++++++++++++++++++++%
\begin{equation}
|\psi_\textrm{MF}\rangle = \prod_{i=1}^{N} \left( \sum_{n=0}^{n_\textrm{max}}\sum_{m=0,1} g^{(i)}_{n,m}|n,m\rangle_i \right)
\label{Ansatz}
\end{equation}
%+++++++++++++++++++++++++++++++++++%
where $n_\textrm{max}$ is an arbitrary maximum occupation number of
bosons in each lattice site \cite{foot2}. 

The $g^{(i)}_{n,m}$ are complex coefficients proportional to the amplitude of finding $n$ bosons
and $m$ fermions in the $i$-th lattice site, and consequently we can impose, without loss of generality,
these coefficients to satisfy $\sum_{n,m}|g^{(i)}_{n,m}|^2=1$.
For the sake of simplicity, we neglect the anticommutation relation of fermionic creation ($f_i$) and annihilation ($f_j^\dagger$) operators in different lattice sites. However Pauli principle applies in each lattice site ($m_i \leq 1$ $\forall i$). Since GA neglects correlations between different sites, this procedure is expected to be safe and is
commonly used within the Gutzwiller approach \cite{fb}.

Inserting $|\psi_\textrm{MF}\rangle$ in the Schr\"odinger equation with the two-species
Fermi-Bose Hamiltonian~(\ref{hamiltonian}), we were able to determine the ground state and
to compute the dynamical evolution of the Fermi-Bose mixture.

\paragraph*{Ground state calculations -}
Employing
% a variational method and using
a standard conjugate-gradient downhill method \cite{recipes}, we
minimize the total energy $\langle \psi_{MF}|H_\textrm{FBH}|\psi_{MF}\rangle$
 with $H_{FBH}$ given by (1) under the constraint of fixed total numbers of fermions
$N_\textrm{f}$ and bosons $N_\textrm{b}$ \cite{foot3}:

\begin{widetext}
%+++++++++++++++++++++++++++++++++++%
\begin{equation}
\langle \psi_\textrm{MF}| H_\textrm{FBH} |\psi_\textrm{MF}\rangle
-\Lambda_\textrm{f} \left(\langle \psi_\textrm{MF}| \sum_i m_i| \psi_\textrm{MF}\rangle-N_\textrm{f}\right)^2
-\Lambda_\textrm{b} \left(\langle \psi_\textrm{MF}| \sum_i n_i| \psi_\textrm{MF}\rangle-N_\textrm{b}\right)^2
~ \rightarrow ~ \textrm{min} ~.
\label{minim}
\end{equation}
%+++++++++++++++++++++++++++++++++++%
\end{widetext}
The numerical procedure is as follows: (i) We minimize
the energy of the mixture (eventually) in presence of smooth trapping potentials and with
non-zero tunneling for bosons and fermions, but assuming vanishing interactions between
bosons and fermions ($U=0$). During the minimization the normalization ($\sum_n \sum_m |g_{n,m}|^2=1$)
should be imposed.
(ii) After this initial minimization, we ramp up adiabatically the
interactions between bosons and fermions using the dynamical Gutzwiller approach
(see below). In this way, we end up with the ground state of the mixture in presence of tunneling $J$,
non-vanishing interactions $U$ and $V$ and eventually in the the presence of a smooth trapping potential.

This two-step procedure is indeed necessary because
in the presence of interactions between bosons and fermions
 finite numbers of bosons and fermions correspond to a saddle point of Eq.~(\ref{minim}),
and no true minimum can be found within direct minimization
%procedure
of the total Hamiltonian \cite{fboptexp}.

\paragraph*{Time-dependent calculations -}
Using the time-dependent variational
%theory
principle
 ($\langle \psi_\textrm{MF} | i \hbar \partial_t - H_\textrm{FBH} (t) |
\psi_\textrm{MF} \rangle \rightarrow \textrm{min}$) with Hamiltonian $H_{BFH}$ given by~(\ref{hamiltonian})
and eventually time-dependent parameters $J_\textrm{f,b}$, $U$, $V$, $\mu_i^\textrm{f,b}$, we end up with
the following dynamical equation for the Gutzwiller coefficients \cite{dynamGA,fboptexp}~:
%+++++++++++++++++++++++++++++++++++%
\begin{eqnarray}
i \hbar \partial_t g^{(i)}_{n,m} & = &
\left[\frac{V}{2}n(n-1)+Unm-\mu_i^\textrm{b}n-\mu_i^\textrm{f}m\right]g_{n,m}^{(i)} \nonumber\\
&&-\left(\Sigma_i^\textrm{b}\right)\sqrt{n} \ g_{n-1,m}^{(i)}-\left(\Sigma_i^\textrm{b}\right)^*\sqrt{n+1} \ g_{n+1,m}^{(i)}
\nonumber \\
&&-\left(\Sigma_i^\textrm{f}\right)g_{n,m-1}^{(i)}-\left(\Sigma_i^\textrm{f}\right)^*g_{n,m+1}^{(i)} \label{evolution}
\end{eqnarray}
%+++++++++++++++++++++++++++++++++++%
where
%+++++++++++++++++++++++++++++++++++%
\begin{eqnarray}
\Sigma_i^\textrm{b} &=& \sum_{\langle i,j \rangle}{J_{\textrm{b}}
\left[ \sum_{n}\sum_{m=0,1}{\sqrt{n+1} \ g_{n,m}^{(j)*} \ g_{n+1,m}^{(j)}} \right]}\label{evolutionbis_1} \\
\Sigma_i^\textrm{f} &=& \sum_{\langle i,j \rangle}{J_{\textrm{f}}
\left[ \sum_{n}{g_{n,0}^{(j)*} \ g_{n,1}^{(j)}} \right] .}
\label{evolutionbis}
\end{eqnarray}
%+++++++++++++++++++++++++++++++++++%
Note that these equations are valid under the hypothesis of neglecting
anticommutation
relations for fermionic operators in different sites.
Equations~(\ref{evolution}-\ref{evolutionbis})
%conserve the total number of fermions $N_\textrm{b}$ and bosons $N_\textrm{b}$ so that no normalization is required.
preserve both normalization of the wave function and the mean particle numbers.

In the following, dynamical Gutzwiller approach will be used for (i)
computing the ground state of the mixture in the presence of interactions
between bosons and fermions (see above) and (ii) to ramp up adiabatically disorder
in the optical lattice potential.

\subsection{Numerical results}
We have considered a 2D optical lattice with $N=10\times 10$ sites to perform
the simulations of the different quantum phases
 %obtained
that appear for type $\overline{II}$ composites in the presence of a very shallow harmonic
trapping potential ($\mu_i^\textrm{b,f} = \omega^\textrm{b,f}\times l(i)^2$, where $l(i)$ is the
distance from site $i$ to the central site in cell size units), with different amplitudes for bosons
and fermions. The harmonic on-site energy simulates shallow magnetic or optical trapping. The confining potential
is experimentally of vital importance in order to see Mott insulator phases, that require commensurate filling
 %is currently used in many experiments on ultracold atoms trapped in optical lattices to avoid losses
\cite{bloch,esslinger,phillips}.
 %modifies
It plays the role of a local chemical potential, and it has been predicted that it modifies some properties
of strongly correlated phases \cite{fillingfactor}.
Additionally, this breaks the equivalence of all lattice sites and makes it more obvious the different
phases that one can achieve (see below).
We first calculate the ground state of the system considering $N_\textrm{b}=60$ bosons,
$N_\textrm{f}=40$ fermions, $J_\textrm{b}/V=J_\textrm{f}/V=0.02$, $U=0$ in the presence of harmonic
traps  characterized by $\omega^\textrm{b}=10^{-7}$ and $\omega^\textrm{f}=5\times 10^{-7}$.

\begin{figure}
\includegraphics[width=1.0\linewidth]{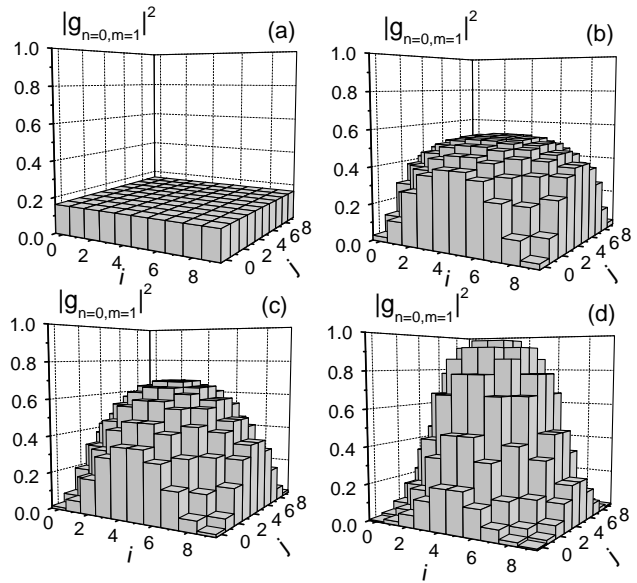}
\caption{Probability of having one fermion and zero boson at each lattice site for
the $N=100$ sites in a Fermi-Bose mixture with $N_\textrm{b}=60$, $N_\textrm{f}=40$, $J_\textrm{b}/V=J_\textrm{f}/V=0.02$
and in the presence of harmonic traps for bosons and fermions characterized by $\omega_\textrm{b}=10^{-7}$ and $\omega_\textrm{f}=5\times 10^{-7}$, respectively.
The interaction between fermions and bosons is (a) $\alpha=0$ [independent bosonic MI and Fermi gas],
(b) $\alpha=0.5$ [Fermi liquid], (c) $\alpha=1$ [ideal Fermi gas] and (d) $\alpha=10$
[fermionic insulator domain].}\label{fig:no_disorder}
\end{figure}

Under these conditions, we find that, as expected, the bosons are well inside the
MI phase with $\tilde n=1$ boson per site \cite{fisher,jaksch}. Due to the very small
values of $\omega^\textrm{f}$ and $\omega^\textrm{b}$ neither the bosons nor the fermions
feel significantly the confining trap as shown in Fig.~\ref{fig:no_disorder}(a).

%-----------------------------------%
%-----------------------------------%

\paragraph*{Non disordered phases -}
Starting with this ground state we adiabatically grow the repulsive interactions between bosons and
fermions, $U$, keeping the repulsion between bosons, $V$, constant, i.e., growing effectively $\alpha$ in order to create the composites.
Once the composites  appear,
 %we find as expected that
the only non-zero probabilities are: (i) $|g^{(i)}_{n=1,m=0}|^2$ to have one boson and zero fermion
(i.e., no composite), or (ii) $|g^{(i)}_{n=0,m=1}|^2$ to have zero boson and one fermion (i.e., one composite).
This proves the formation of type $\overline{II}$ composites \cite{foot4}.
In Fig.~\ref{fig:no_disorder}(b), we show the probability of having one composite $|g^{(i)}_{n=0,m=1}|^2$ (i.e., one
fermion and zero boson) at each lattice site for $\alpha=0.5$, which corresponds to repulsive interactions
between composites $K_\textrm{eff}=d_\textrm{eff}=1.6 \times 10^{-3}$. Due to the important value of the
composite tunneling $d_\textrm{eff}$, the ground state is delocalized and corresponds to a (non-ideal)
{\it Fermi liquid}.

Increasing further the fermion-boson interaction parameter, $\alpha$, the system reaches the point where the
interactions between composites are negligible corresponding to the region of an ideal {\it Fermi gas} phase
($\alpha \simeq 1$). Fig.~\ref{fig:no_disorder}(c) displays the probability of having a composite in each lattice
site in the case where the interactions between composites exactly vanish, i.e., $\alpha=1$. Increasing again
the interaction parameter $\alpha$, one reaches for $\alpha>1$ the region where the interactions between composites are
attractive ($K_\textrm{eff}<0$). In this region, composite {\it fermionic insulator domains} are predicted.
Due to the attractive interactions, the probability of having composite fermions in the center of the
trap increases reaching nearly one for high enough effective attractive interactions as shown in Fig.~\ref{fig:no_disorder}(d).

It is also worth noticing that the energies involving the composites are at least three orders of magnitude
smaller than the corresponding energies for bosons and fermions ($J_\textrm{eff}, K_\textrm{eff} \ll J, U, V$).
As a consequence, the effect of inhomogeneities is much larger for composites than it is for bare bosons and
fermions. This is exemplified in Fig.~\ref{fig:no_disorder}. For no interaction between bosons and fermions
($\alpha=0$), the bare particles are not significantly affected by the harmonic trap on the $10 \times 10$
lattice (see Fig.~\ref{fig:no_disorder}(a)). On the contrary, as soon as composites $\overline{II}$ are created,
 the harmonic trapping clearly reflects in inhomogeneous population of the lattice sites
(see Fig.~\ref{fig:no_disorder}(b)-(d)). Another important consequence is that large time scales are necessary
in time evolution processes in order to fulfill the adiabaticity condition.

\paragraph*{Disordered phases -}
We now consider disordered optical lattices for the bosons. The on-site energy $\mu_i^\textrm{b}$
is assumed to be random with time-dependent standard deviation
$\sqrt{\langle (\tilde{\mu}_i^{\textrm{b}})^2 \rangle}=\Delta(t)$ and independent from site to site.
For this, we create type $\overline{II}$ composites in different regimes (this is controlled by
the value of $\alpha$ as shown before) and we slowly ramp up the disorder from $0$ to its final
value $\Delta$.

\begin{figure}
\includegraphics[width=0.9\linewidth]{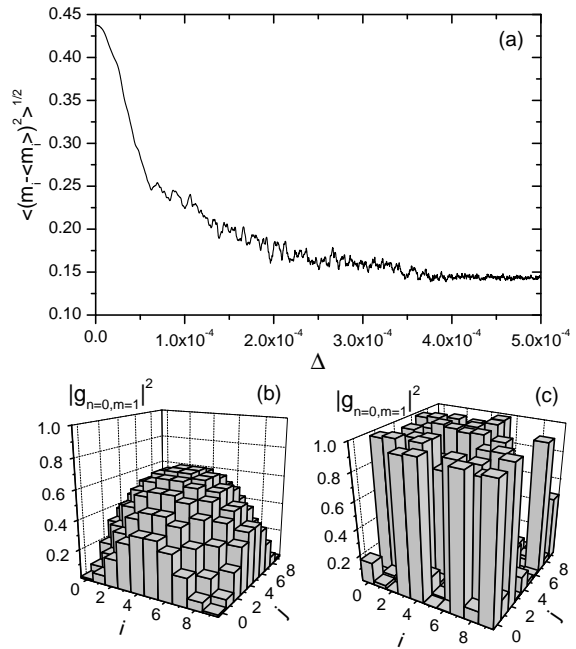}
\caption{Dynamical crossover from the Fermi gas to the Fermi glass phases. The parameters are the same as in
Fig.~\ref{fig:no_disorder}(c). (a) Variance of the number of fermions per lattice site as a function of the amplitude of the disorder $\Delta$. (b) Probability of having one composite (one fermion and zero boson) at each lattice site for the M sites
in the absence of disorder and (c) after ramping up adiabatically diagonal disorder with amplitude $\Delta=5\times 10^{-4}$.}\label{fig:dis_fermi_liquid}
\end{figure}

Let us first consider a {\it Fermi gas} in absence of
disorder (see Fig.~\ref{fig:dis_fermi_liquid}(b)). Because of effective tunneling, $d_{ij}$,
the composite fermions are delocalized although confined near the center of the effective
harmonic potential ($(\omega_\textrm{f}-\omega_{b})\times l(i)^2$). In particular, the population
of each lattice site fluctuates around  $\langle m_i \rangle \simeq 0.4$
%{\bf ADD THE VALUE FROM THE SIMULATIONS}
with
$\sqrt{\langle (m_i - \langle m_i \rangle)^2 \rangle} \simeq 0.43$. While slowly increasing
the amplitude of disorder, the composite fermions become more and more localized in the lattice
sites to form a {\it Fermi glass}. Indeed, Fig.~\ref{fig:dis_fermi_liquid}(a) shows that the
fluctuations in composite number are significantly reduced as the amplitude of the disorder increases. For $\Delta=5 \times 10^{-4}$,
 the composite fermions are pinned in random sites as shown in Fig.~\ref{fig:dis_fermi_liquid}(c).
As expected, the $N_\textrm{f}$ composite fermions populate the $N_\textrm{f}$ sites with
minimal $\tilde{\mu}_i^\textrm{b}$.

It should be noted that in the absence of interactions between bosons and fermions
(i.e., when the composites are not formed), no effect of disorder is observed. This again
shows the formation of composites with typical energies significantly smaller than those of bare particles.

%-----------------------------------%

%-----------------------------------%

\begin{figure}
\includegraphics[width=0.9\linewidth]{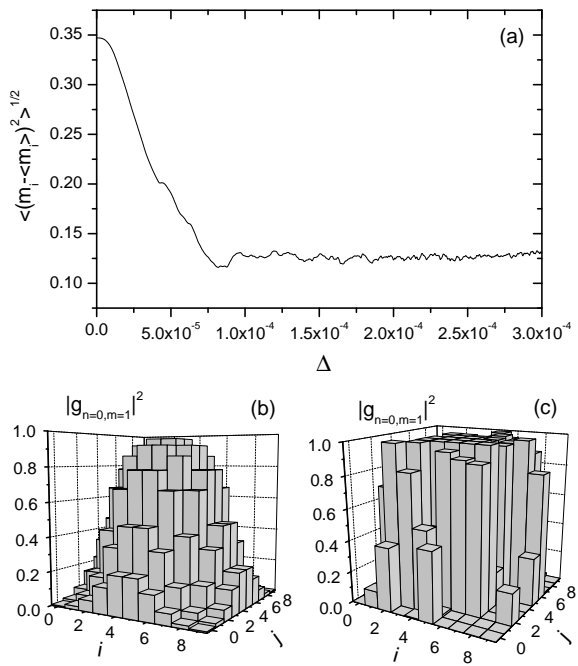}
\caption{Dynamical crossover from the fermionic domain insulator to a
disordered insulating phase. The parameters correspond to
Fig.~\ref{fig:no_disorder}(d). (a) Variance of the number of fermions per lattice site as
a function of the amplitude of the disorder. (b) Probability of having one composite (one fermion and zero boson) in
 each lattice site in the absence of disorder and (c) after ramping up adiabatically
the disorder with amplitude $\Delta=3\times 10^{-4}$.}\label{fig:dis_domains}
\end{figure}

We now consider the {\it Fermi insulator domain} phase (see Fig.~\ref{fig:dis_domains}(b))
with slowly increasing disorder. In the {\it Fermi insulator domain} (in the absence of disorder),
the composite fermions are pinned in the central part of the confining potential. In addition,
there is a ring of delocalized fermions and this gives finite fluctuations on the atom number
per site ($\sqrt{\langle (m_i - \langle m_i \rangle)^2 \rangle} \simeq 0.35$). As shown
in Fig.~\ref{fig:dis_domains}(a), while ramping up the amplitude of disorder, the fluctuations
decrease fast and reach ($\sqrt{\langle (m_i - \langle m_i \rangle)^2 \rangle} \simeq 0.13$
for $\Delta>10^{-4}$. This indicates that the composite fermions are pinned in different lattice sites.
This is confirmed in Fig.~\ref{fig:dis_domains}(c) where we plot the population of the composites fermions in each
lattice site for $\Delta=5\times 10^{-4}$. Contrary to what happens for the transition from
{\it Fermi gas} to {\it Fermi glass}, the composites mostly populate the central part of the confining
potential. The reason for that is twofold. First, with our parameters, the attractive interaction between
composites is of the order of $K \simeq -1.4\times 10^{-3}$ and can compete with disorder
$\Delta = 3\times 10^{-4}$. This explains the central insulator domain. Second, because
tunneling is small ($d\simeq 8\times 10^{-5}$) and because disorder breaks the symmetry of
lattice sites in the ring around the domain, the atoms in this region get pinned. The populated
 sites match the lowest $\tilde{\mu}_i^\textrm{b}$.
%-----------------------------------%
%-----------------------------------%
%%%%%%%%%%%%%%%%%%%%%%%%%%%%%%%%%%%%%%%%%%%%%%%%%%%%%%%%%%%%
\section{Spin glasses}

In this chapter we discuss in  more detail the possible realization of the Edwards-Anderson
spin glass Fermi-Bose mixtures as discussed in section~\ref{FBhc}.
Strictly speaking, since the system is quantum it allows for realization
of  fermionic spin glass \cite{oppermann}. 
The main goal of such investigation is to study the nature of the
spin glass ordering and to compare the predictions of the M\'ezard-Parisi and "droplet" pictures.

Although we work along the lines of the original papers \cite{parisi}, it is necessary
to reformulate the standard M\'ezard-Parisi
mean field description of our system. 
The main difference appears because the Ising spins are coded as presence or absence of
a composite at a given site. This leads to a fixed magnetization due to the fixed number of particles
in the system.
For this we repeat very shortly the Sherrington-Kirkpatrick calculations \cite{sherr} here,
adapted to our case.

\label{sec:spinglass}
\subsection{Edwards-Anderson model for composite fermions}
The spin glass limit obtained in section~\ref{FBhc} with large disorder derives 
of the composite fermionic model (\ref{Heffinhom}) with vanishing hopping due to
strong site-to-site energy fluctuations and n.n.-interactions, $\Kij$. By appropriate
choice of $\Deltaij^\textrm{b}$, $\Kij$ fluctuate around mean zero 
with random positive and negative values (see Fig. \ref{fig:sitesB}(b)). 
Replacing the
composite number operators with a classical
Ising spin variable $s_i:=2M_i-1=\pm 1$, one ends up with the Hamiltonian:
%+++++++++++++++++++++++++++++++++++%
\begin{equation}
H_\textrm{E-A}=\frac{1}{4}\sum_{\left\langle ij\right\rangle }
K_{ij}s_is_j + \frac{1}{2}\sum_{i}\overline\mu_i s_i ,
\label{HamiltonianSG}
\end{equation}
%+++++++++++++++++++++++++++++++++++%
It describes an (fermionic) Ising spin glass \cite{oppermann}, which differs from the Edwards-Anderson model
\cite{binder,parisi} in that it has
an additional random magnetic field $\overline{\mu}_i$ and moreover has to satisfy the constraint of
fixed magnetization value, $m=2N_\textrm{f}/N-1$, as the number of fermions in the underlying BFH-model is conserved.
It however shares the basic characteristics with the Edwards-Anderson model as being a
spin Hamiltonian with random spin exchange terms $\Kij$. In particular,
this provides bond frustration, which
in this model is essential for the appearance of a spin glassy phase. 
The experimental study of this limit thus could present a way to address various open questions of spin glass physics concerning the nature and the ordering of its ground- and possibly metastable states
(the M\'ezard-Parisi picture \cite{parisi} versus the ``droplet'' picture \cite{huse,stein}),
broken symmetry and dynamics in classical (in absence of hopping) and quantum (with small, but
nevertheless present hopping) spin glasses \cite{sachdev,georges}.

For sufficiently large systems, Eq.~(\ref{HamiltonianSG}) is well approximated by assuming \Kij\ and $\overline \mu_i$ to be
independent random variables with Gaussian distribution, with mean $0$ and $H$, respectively and variances $K/\sqrt{N}$
and $h$, respectively \cite{foot5}. This approximation will be used in the following calculations.

Before employing the mean field approach for Edwards-Anderson-like models in section \ref{sksubsec} for
the Hamiltonian
(\ref{HamiltonianSG}), a very basic outline of the different phases of the short-range
Ising models with bond frustration
is given and the two competing physical pictures for the spin glass phase are briefly summarized
in this section.

The experimental observations have led to the identification of three equilibrium phases,
which are characterized by two order parameters (for zero external magnetic field): 
\ensm{\M:=\overline{\avt{\si}}} is the magnetization, i.e., the order parameter for magnetic ordering,
and \ensm{\QEA:=\overline{\avt{\si}^2}} is
the Edwards-Anderson order parameter for spin glass ordering. Here, $\avt{\cdot}$ denotes the Gibbs ensemble
average and
$\overline{\hspace{0.1cm}\cdot\hspace{0.1cm}}$ the disorder average.
The three phases are:
(i) an unordered paramagnetic phase, with
$\M=0$ and \ensm{\QEA=0}; (ii) an ordered spin glass phase with $\M=0$ and \ensm{\QEA\neq0}
that is separated from the paramagnetic
phase by a second order phase transition \cite{foot6};
and (iii) dependent on the mean value of \Kij, an ordered ferromagnetic phase with $\M\neq0$
and $\QEA=0$. 
It should be pointed out that there are additional questions - different from, but of course connected to the ones discussed
in the following - about the nature of the equilibrium spin glass state that stem from the intrinsic problems that are associated
in this system with separating equilibrium from non equilibrium effects such as metastability, hysteresis and others
(see \cite{refrigier} and references therein).

\subsection{M\'ezard-Parisi picture}

The {\it M\'ezard-Parisi} (MP) picture is fundamentally guided by the results of the mean-field theory.
On the level of statistical physics, the Gibbs equilibrium
distribution of the spin system in the MP-picture at temperature $T$ and for a particular
disorder configuration $K$ can be written as a unique convex combination of
infinitely many pure equilibrium state distributions \cite{vanenter,stein},
%+++++++++++++++++++++++++++%
\begin{equation}\label{rsb1}\rhokt=\suma\wakt\rhoakt,\quad\mbox{with}\quad\suma\wakt=1
\end{equation}
%+++++++++++++++++++++++++++%
where the overlap between two pure states is defined as
%+++++++++++++++++++++++++++%
\begin{equation}\label{rsb3}\Qab \equiv \Om^{-1}\sumi\avtka{\si}\avtkb{\si}
\end{equation}
%+++++++++++++++++++++++++++%
where \Om\ denotes the size of the system. The mean-field version of \Qab\ emerges naturally
from the calculation in the next section and motivates definition (\ref{rsb3}).

In the MP-picture, the spin glass transition is interpreted akin to the transition
from an Ising paramagnet to a ferromagnet. There, the Gibbs distribution
is written as a sum of only two pure states, corresponding
to the two possible fully spin-polarised ferromagnetic ground states. As the temperature of the system decreases,
the $Z_2$ symmetry of the system is broken, and a phase
transition to a ferromagnetic phase occurs, whose equilibrium properties
are not described by the Gibbs state, but by the relevant pure state
distribution alone \cite{parisi}. Analogously, the spin glass transition is characterized
by the breaking of the infinite index symmetry, called replica symmetry breaking in the mean-field case,
by which one pure state
distribution \rhoakt\ is chosen and alone describes the low-temperature
properties of the system \cite{parisi}. However, unlike the Ising ferromagnet, the pure
states of the spin glass are not related to each other by a symmetry of the
Hamiltonian, but rather by an accidental, infinite degeneracy of the ground
state caused by the randomness of the bonds and the frustration effects.
This picture can be interpreted as the system getting frozen into one particular
state out of infinitely many different ground- or
metastable states of the system. These states are all taken to be separated by free energy
barriers, whose height either diverges with the system
size or it is finite but still so large that the decay into a ''true'' ground state does not
occur on observable timescales. Thus, fluctuations around one of these ground states
can only sample excited states within one particular free energy valley.
Consequently, \QEA\ in the spin glass phase must be redefined as \ensm{\QEA:=\Qaa}, the self-overlap of the
state, whereas it remains unchanged for the paramagnetic phase.

\subsection{de Almeida-Thouless plane}

Based on the results of mean-field theory, one of the predictions of the MP-picture concerns
the order of the infinitely many spin glass ground states, which is ultrametric \cite{parisi,foot7},
as can be seen from the joint probability distribution of
three different ground state overlaps,
$P_K(Q_{12},Q_{13},Q_{23})$. Upon choosing independently
three pure states $1$, $2$ and $3$ from the decomposition (\ref{rsb1}),
one should find that with probability $1/4$, $Q_{12}=Q_{13}=Q_{23}$ and with
probability $3/4$ two of the overlaps are equal and smaller than the
third. Ultrametricity then follows from the canonical
distance function $D_{\alph\beta}=\QEA-\Qab$.
The mean-field theory, both with and without a magnetic field, predicts the existence of a plane in the space of the
Hamiltonian parameters,
called de Almeida-Thouless (dAT) plane \cite{at}, below which the naive ansatz for the spin glass phase becomes invalid
and the system is characterized by the transition to this ultrametrically ordered infinite manifold of ground states.
It should be pointed out that the clear occurrence of such a dAT-plane in the finite range Edwards-Anderson model would be an important indicator for the validity
of the MP-picture in these systems. As we discuss in section \ref{droplet}, this conclusion has to be drawn with great care.

\subsection{"Droplet" model} \label{droplet}

The very
applicability of the MP-model for finite-range systems is however still unproven. It is both
challenged by a rival theory, the
so-called {\it droplet} model \cite{huse}, as well as by mathematical
analysis (cf. \cite{stein} and references therein) that questions the validity of transferring a
picture developed
for the infinite-range mean-field case to the short-range model.
Being a phenomenological theory based on scaling
arguments and numerical results, the droplet model describes the ordered spin glass phase below the
transition as one of just
two possible pure states, connected by spin-flip symmetry, analogous to the ferromagnet mentioned above.
Consequently there can be no infinite hierarchy
of any kind, and thereby no ultrametricity. Excitations over the ground state are regions with a
fractal boundary - the droplets - in which the spins are in the
configuration of the opposite ground-state.
The free energy of droplets of diameter $L$ is taken to scale as $\sim L^{\theta}$, with $\theta<0$
at and below the critical dimension,
which is generally taken to be two. So three is the only physical dimension where the spin glass transition is stable
with nonzero transition temperature, with $\theta\sim0.2$ in this case. The free energy barriers for the creation and annihilation
of a droplet scales in 3D as $\sim L^{\psi}$, with $\theta\leq\psi\leq2$.

Although there can be no dAT-plane in the strict sense in the droplet-model, for an external magnetic field the system
can be kept from equilibration on experimental timescales for parameters below a line that scales just like the dAT-line.
This phenomenon might mimic the effects of the replica symmetry-breaking in the MP-picture (see \cite{huse} for
further details).

\subsection{Replica-symmetric solution for fixed magnetization}
\label{sksubsec}

This section serves to show that the mean-field version of the effective Hamiltonian
(\ref{HamiltonianSG}) with random magnetic field and magnetization constraint
exhibits replica symmetry breaking just as for the pure Edwards-Anderson model, and
would therefore be a candidate to examine the validity of the MP or droplet-picture in a realistic short-ranged spin-glass model.
Following Sherrington and Kirkpatrick (SK) \cite{parisi}, the mean-field model is given by:
%++++++++++++++++++++++++++++++++++%
\begin{equation}\label{hsk} H_{SK}=
\frac{1}{4}\psum \Kij\si\sj+\frac{1}{2}
\sumi \overline{\mu}_i\si
\end{equation}
%++++++++++++++++++++++++++++++++++%
where the round brackets $(\cdot,\cdot)$
are generally used to denote sums over all pairs of different indices.
This model differs from Eq.~(\ref{HamiltonianSG}) by the long-range spin exchange. 
As the mean of $\overline{\mu}_i$, $H$, is generally
nonzero this model will not exhibit a phase transition, which however is not a concern, as the
number of (quasi-)ground states will be the quantity of interest. Following the
analysis of SK, we aim at finding the free energy, ground state overlap and magnetization constraint. 
Then we will use the de Almeida and Thouless approach \cite{parisi} to show that the obtained solution is unstable
in a certain parameter region, that lies below the so-called dAT-plane of stability.
The type of instability that emerges is then well known to require the replica symmetry breaking solution of 
Parisi \cite{parisi}.

% To deal with the random variables in (\ref{hsk}),
% the replica-trick is employed. As the disorder is quenched (static on
% experimental timescales), one cannot average it directly
% in the partition function as would be done for annealed disorder,
% i.e., $\overline{Z}$, but must rather average the free energy density,
% \ensm{\overline{f}=-\beta\overline{\ln Z}}. To execute the average, the formula
% \ensm{\ln x=\lim_{n\rightarrow0}(x^n-1)/n} is employed, where the $n$ copies of the system are the replicas.
% The average is calculated for integer $n$ and a finite number of spins $N$ and then analytically continued to zero before
% taking the thermodynamic limit. Explicitly, \ensm{\overline{Z^n}}
% is given by:
% %+++++++++++++++++++++%
% \begin{equation}\label{rsb9}\overline{Z^n}=
% \sum_{\{\sia=\pm1\}}\exp\lsb-\beta \overline{H[\sia,n]}\rsb
% \end{equation}
% %+++++++++++++++++++++%
% where $\overline{H_{SK}[\sia,n]}$ is the sum of $n$ spin Hamiltonians (\ref{hsk}), averaged over the
% Gaussian disorder, with Greek indices now numbering the replicas.

As the disorder is quenched (static on
experimental timescales), one cannot average directly over disorder
in the partition function as would be done for annealed disorder, 
but one must rather average the free energy density,
\ensm{\overline{f}=-\beta\overline{\ln Z}} using the 'replica trick':
We form $n$ identical copies of the system (the replicas) and 
the average is calculated for an integer $n$ and a finite number of spins $N$. 
Then, using the general formula \ensm{\ln x=\lim_{n\rightarrow0}(x^n-1)/n}, 
\ensm{\overline{\ln Z}} is obtained  from the analytic continuation of \ensm{\overline{Z^n}}
for \ensm{n \rightarrow 0}. Finally, we take the thermodynamic limit \ensm{N \rightarrow \infty}. 
Explicitly, \ensm{\overline{Z^n}}
is given by:
%+++++++++++++++++++++%
\begin{equation}\label{rsb9}\overline{Z^n}=
\sum_{\{\sia=\pm1\}}\exp\lsb-\beta \overline{H[\sia,n]}\rsb
\end{equation}
%+++++++++++++++++++++%
where $\overline{H_{SK}[\sia,n]}$ is the sum of $n$ independent and indentical 
spin Hamiltonians (\ref{hsk}), averaged over the
Gaussian disorder, with Greek indices now numbering the $n$ replicas.

Executing the average over the Gaussian distributions for \Kij\ and $\overline{\mu}_i$ leads
to coupling between spin-spin-interactions of different replicas.
As the mean-field approach means that the double sum over the site indices in (\ref{hsk}) can be simplified into a square
using \ensm{(\sia)^2=1}, one finds:
\begin{widetext}
%+++++++++++++++++++++++++%
\begin{eqnarray}\label{sk4}\lefteqn{\overline{\fNn}=-(Nn\beta)^{-1}\lgb e^{
Nn(\beta K)^2/4}e^{-n^2(\beta K)^2/2+n(\beta h)^2/2}
\Nnspintr\exp\lsb\frac{N(\beta K)^2}{2}\sumalb\lrb\sumi
\frac{\sia\sib}{N}\rrb^2\right.\right.}\hspace{8cm}\nonumber
\\& &\left.\left.+(\beta h)^2\sumalb\sumi\sia
\sib-\beta H\sum_{\alph}\sumi\sia\rsb-1\rgb
\end{eqnarray}

%\begin{equation}\label{sk4}\overline{\fNn}=-(Nn\beta)^{-1}\lgb e^{
%Nn(\beta K)^2/4}e^{-n^2(\beta K)^2/2+n(\beta h)^2/2}
%\Nnspintr\exp\lsb\frac{N(\beta K)^2}{2}\sumalb\lrb\sumi
%\frac{\sia\sib}{N}\rrb^2 +(\beta h)^2\sumalb\sumi\sia
%\sib-\beta H\sum_{\alph}\sumi\sia\rsb-1\rgb
%\end{equation}
%+++++++++++++++++++++++++%
\end{widetext}
where the prefactor \ensm{e^{-n^2(\beta K)^2/2+
n(\beta h)^2n/2}} becomes irrelevant in the limit
$n\rightarrow 0$ and is subsequently dropped. 
As in the standard procedure, the square of the operator sum
\sumi\sia\sib\ is decoupled by introducing auxiliary operators
\qab\ via a Hubbard-Stratonovitch transformation \cite{parisi}.
\begin{widetext}
%++++++++++++++++++++++++++++%
%\begin{equation}\label{sk5}\overline{\fNn}=-(Nn\beta)^{-1}\lgb
%e^{Nn(\beta K)^2/4}\infint\lsb\prod_{\alph<\beta}
%d\qab\lrb\frac{N}{2\pi}\rrb^{\frac{1}{2}}\beta
%K\rsb\exp\lsb-\frac{N(\beta K)^2}{2}\sumalb\qab^2+N\ln
%\lrb\nspintr\exp\lsb L(\qab)\rsb\rrb\rsb-1\rgb
%\end{equation}

\begin{eqnarray}\label{sk5}\lefteqn{\overline{\fNn}=-(Nn\beta)^{-1}\lgb
e^{Nn(\beta K)^2/4}\infint\lsb\prod_{\alph<\beta}
d\qab\lrb\frac{N}{2\pi}\rrb^{\frac{1}{2}}\beta
K\rsb\right.}\hspace{2.0cm}\nonumber\\& &\left.\times
\exp\lsb-\frac{N(\beta K)^2}{2}\sumalb\qab^2+N\ln
\lrb\nspintr\exp\lsb L(\qab)\rsb\rrb\rsb-1\rgb
\end{eqnarray}
%++++++++++++++++++++++++++++%
\end{widetext}
where the functional $L(\qab)$ is:
%++++++++++++++++++++%
\begin{equation}\label{sk6}L(\qab)=\beta^2\sumalb\lrb K^2\qab+
h^2\rrb\sa\sbb-\beta H\suma\sa
\end{equation}
%++++++++++++++++++++%
and the configuration sum of \ensm{\exp\lsb
L(\qab)\rsb} now only goes over the $n$ spins $\sa$ in
$L(\qab)$, the HS-transformation having made it possible
to decouple the configuration sum over $Nn$ spins
in (\ref{sk4}) into a $N$-fold product of $n$-spin sums.
Assuming that the thermodynamic limit ($N\rightarrow\infty$)
can be taken before $n\rightarrow 0$, i.e., that the
usual limiting process can be inverted, then Eq.~(\ref{sk5})
can be evaluated by the method of steepest descent,
as the exponent is proportional to $N$. According to
this method, the free energy per spin in the thermodynamic
limit is the maximum of the $\qab$-dependent function
in the exponent:
\begin{widetext}
%++++++++++++++++++++%
\begin{equation}\label{sk6a}-\beta \overline{f}=\lim_{n\rightarrow 0}
\max\lgb\frac{(\beta K)^2}{4}\lrb1-\frac{1}{n}\sumpab\qab^2
\rrb+\frac{1}{n}\ln\lrb\nspintr\exp\lsb L(\qab)\rsb\rrb\rgb
\end{equation}
%++++++++++++++++++++%
\end{widetext}
(with $\overline{f}:=\lim_{n\rightarrow 0} \overline{f^n}$) with the self-consistency condition:
%++++++++++++++++++++%
\begin{equation}\label{sk7}\frac{\partial \overline{f}}{\partial \qab}
=0\quad\Leftrightarrow\quad\qab=\langle\sa\sbb\rangle_L
\end{equation}
%++++++++++++++++++++%
and the magnetization:
%++++++++++++++++++++%
\begin{equation}\label{sk7a} m=-\frac{1}{\beta}\frac{\partial\overline{f}}{\partial H}=\langle\sa\rangle_L = 2N_F/N-1
\end{equation}
%++++++++++++++++++++%
where the average $\langle(\cdot)\rangle_L$ is defined as:
%++++++++++++++++++++%
\begin{equation}
\langle(\cdot)\rangle_L=\lim_{n\rightarrow 0}\frac
{\nspintr(\cdot)\exp\lsb L(\qab)\rsb}{\nspintr\exp\lsb
L(\qab)\rsb}
\end{equation}
%++++++++++++++++++++%
The mean-field approach has allowed a decoupling of the spins,
and a reduction of the problem to a single-site model with ''Hamiltonian''
$L[\qab]$. For this new problem, the overlap parameter emerges naturally,
albeit in a self-consistent manner. To push the calculation further 
some assumption for $\qab$ has to be made. Naively, from the
requirement that the result should be independent of the replica-indices,  the most natural choice for
$q$ is to consider all identical
overlaps between the replicas, $\qab=q$, which is the
SK-ansatz. Thus the double sum over the replicas \ensm{\sumpab\sa\sbb} in (\ref{sk6})
can be written as a square, keeping $(\sa)^2=1$ in mind. Another Hubbard-Stratonovitch-transformation with auxiliary
variable $z$ then decouples
the square and yields an expression for the free energy density, which has to be evaluated
self-consistently and in
which the limit $n\rightarrow 0$ can be easily calculated:
\begin{widetext}
%+++++++++++++++++++++%
\begin{eqnarray}\label{sk9}\lefteqn{-\beta \overline{f}_{SK}=\lim_{n\rightarrow 0}
\lgb\frac{(\beta K)^2}{4}\lrb1-(n-1)q^2
\rrb+\frac{1}{n}\ln\lrb\frac{1}{\sqrt{2\pi}}\infint dz \exp\lsb-
\frac{z^2}{2}\rsb\right.\right.}\hspace{1cm}\nonumber\\& &\left.
\left.\times\lrb\exp\lsb-\frac{\beta^2(K^2q+h^2)}{2}\rsb
\sum_{s=\pm 1}\exp\lsb\beta\sqrt{K^2q+h^2}s-\beta H s\rsb\rrb^n
\rrb\rgb\end{eqnarray}
%+++++++++++++++++++++%
%+++++++++++++++++++++%
\begin{equation}\label{sk10}\Rightarrow \beta \overline{f}_{SK}=
\frac{(\beta K)^2}{4}\lrb1-q
\rrb^2-\frac{(\beta h)^2}{2}+\lrb\frac{1}{2\pi}\rrb^{\frac{1}{2}}
\infint dz e^{-\frac{z^2}{2}}\ln\lrb2\cosh(\Az)\rrb
\end{equation}
%+++++++++++++++++++++%
\end{widetext}
with \ensm{\Az:=\beta\sqrt{K^2q+h^2}-\beta H}.
The overlap (\ref{sk7}) and the magnetization constraint (\ref{sk7a}) can also be evaluated in the same way:
%+++++++++++++++++++++%
\begin{equation}\label{sk11}
q = \frac{1}{\sqrt{2\pi}}
\infint dz e^{-\frac{z^2}{2}}\tanh^2(\Az)
\end{equation}
%\begin{eqnarray}\underline{q} & \stackrel{(\ref{sk7})}{=} &
%\lim_{n\rightarrow 0}\frac{(2\pi)^{-\frac{1}{2}}\infint
%dz e^{-\frac{z^2}{2}}\sinh^2(\Az)\cosh^{n-2}(\Az)}{\underbrace{(2\pi)^
%{-\frac{1}{2}}\infint dz e^{-\frac{z^2}{2}}\cosh^n(\Az)}_{\rightarrow 1}}
%\nonumber \\ \label{sk11}& = & \underline{\lrb\frac{1}{2\pi}\rrb^{\frac{1}{2}}
%\infint dz e^{-\frac{z^2}{2}}\tanh^2(\Az)}
%\end{eqnarray}
%+++++++++++++++++++++%
%+++++++++++++++++++++%
\begin{equation}\label{sk12}
m=\frac{1}{\sqrt{2\pi}} \infint
dz e^{-\frac{z^2}{2}}\tanh(\Az) = 2N_F/N-1 ~.
\end{equation}
%+++++++++++++++++++++%
A well known problem with the SK-ansatz for \qab\ is that it yields negative entropy for low temperatures
and thus becomes unphysical. This is due to a fundamental technical problem with the replica trick:
For the method of steepest descent to be valid, the SK-solution
must be a maximum of the exponent in Eq.~(\ref{sk5}) and must {\it stay} a maximum
as the replica-limit $n\rightarrow0$ is taken. But there is {\it no} unique
way of choosing the zero-dimensional limit of the matrix \qab. The SK-solution just corresponds to one
possible choice for this limit. Thus, the question arises whether the
SK-solution for the free energy is still a good, i.e., maximal choice in
the replica limit.

To answer this question, one proceeds analogous to the de Almeida and Thouless procedure \cite{at} 
to analyze the fluctuations around the
SK-solution, while taking the magnetization constraint into account (see appendix~\ref{atsubsec} for details). 
Developing (\ref{sk6a}) and (\ref{sk7a}) to second and first order respectively around $\qab=q$ one finds
that in the replica limit $n\rightarrow 0$ there is an eigenvalue $\lambda_2$ of the matrix
$\partial^2\overline{f}/\partial\qab\partial q_{\gamma\delta}$ that can have both negative values and respects the constraint,
yielding the condition:
\begin{equation}\label{at90}\lambda_2=\frac{1}{(\beta K)^2}-\frac{1}{\sqrt{2\pi}} \infint dz e^{-\frac{z^2}{2}}\sech^4(\Az)>0
\end{equation}
which is violated for low enough $T/K$, $H/K$ and $h/K$.
The plane in parameter space below which this happens is the so-called dAT-surface.
This instability is rectified by a much more involved ansatz for $\qab$ that breaks the
symmetry of the replicas and leads to the phenomena described in section A.

%%%%%%%%%%%%%%%%%%%%%%%%%%%%%%%%%%%%%%%%%%%%%%%%%%%%%%%%%%%%%
\section{Experimental insights}
\label{sec:experiments}

Experimental creation and detection of the phenomena discussed in this paper poses
a major problem, and deserves itself a lot of creative thinking and separate publications.
In this section we will just sketch, what are in our opinion the most obvious
ways of addressing these problems. In this section we do not address the questions concerning
experimental realization  of ultracold FB mixtures and composite fermions - these questions
are discussed in Ref.~\cite{fb}.

The first question thus to be addressed is what are the best ways to create quenched disorder in a controlled way. Roth and Burnett \cite{keith}
and we \cite{boseglass} have suggested that the use of pseudo-random disorder induced by
non-commensurate optical lattices should work as well as the use of the genuine random lattices. Indeed the
latter can be only (so far) achieved using speckle radiation,
i.e. disorder correlation
length of order of few microns! If  we work with systems of size of mm's, such disorder
would definitely be enough to induce localization in 1D, or 2D. Unfortunately, the
size of the systems in question is  typically is of order of hundred microns, and that is
one of the reasons, why it is difficult to observe Anderson localization with BEC's
\cite{inguscio,aspect}.

The analysis performed by us  in this context implies clearly that it will be much
easier to achieve the desired properties of the disorder using pseudo-disordered, i.e.,
overlapped  incommensurable optical lattices \cite{arlt}. One should also stress the
equally promising look to the proposals formulated recently to use the optical
tweezers techniques \cite{privat3} and a random distribution of impurity atoms pinned 
in different lattice sites \cite{privat1}.
 
 Completely another variety of problems is related to the detection 
of the quantum phases discussed in this paper. Below we list basic mehtods 
that have  been already successfully applied to ultracold atomic gases
in optical lattices.

\begin{itemize}

\item {\it Imaging of the atomic cloud after ballistic expansion.} This 
(perhaps the most standard method)  
has been used in   Ref. \cite{bloch} to distinguish the bosonic 
SF phase from the MI phase. 
It allows for measurement of the quasi-momentum distribution of atoms obtained after 
initial expansion caused by interactions \cite{fermisur}, i.e., it detects first order coherence (interference pattern), 
present for instance in the SF phase.  

\item {\it Monitoring the density profile.}Using phase contrast imaging \cite{ketterlehulet} it is possible to perform a direct and non-destructive observation of the spatial distribution of the condensate {\it in situ}. This kind of measurement allows the direct observation of superfluidity \cite{ketterlem2} and can be therefore applied to characterize the fluid and superfluid phases. 

\item{\it Tilting or acceleration of the lattice.} This method was used in Ref. \cite{blochnat2}
to detect the gap in the MI phase. It allows, in principle, to distinguish gapless 
from gapped phases, provided the continuum of low energy states can be achieved via tilting. Fluidity, 
superfluidity, and in general extended, non-localized excitations should allow to detect Bloch oscillations 
\cite{blochosc}.

\item{\it Absorption of energy via modulation  of the lattice.} This method was also designed
to detect the gap in the MI phase \cite{esslinger}. Similarly as tilting, it provides a way of
probing excitations in the systems. 

\item{\it Bragg spectroscopy.} This method, one of the first proposed \cite{stenger}, 
is also a way of probing certain kind of excitations
in the system \cite{orsayelongated}.

\item{\it Cooper pair spectroscopy.} This method is particularly useful to detect Fermi superfluids \cite{revBCS}. Its 
theoretical aspects are discussed in Refs.~\cite{paivi}, while for experiments see Ref.~\cite{grimm}. 

\item{\it Trap shaking and nonlinear dynamics.} Yet another way to probe excitations could 
correspond to analysis of the 
response of the system upon sudden shaking of the trap \cite{lukin}. 

\item{\it Observations of vortices, solitons etc.} This method provides a direct way to detect superfluidity
(for vortices in Bose superfluid see for instance Ref.~\cite{jean}, for vortices in Fermi superfluid see
\cite{ketterlelast}). 

\item{\it Spatial quantum noise interferometry}. The last,  but not the least method discussed here 
 allows for practically direct 
measurement of the density-density correlation, and second order coherence. It has been proposed 
in Ref. \cite{altman}
(see also \cite{kazik}), and used with great success to detect the bosonic Mott insulator
\cite{follig} and the Fermi superfluid \cite{greinernat}. It has capabilities of 
detecting and measuring relevant properties of various phases and 
structures ranging from supersolids, charge-density wave phases, and even Luttinger 
liquids in 1D. In the 
Fourier frequency and momentum domain it corresponds to measurements of the 
dynamic structure factor (for a discussion 
in the context of cold gases see Ref. \cite{rothstruc}.

\end{itemize}

There are of course more methods than the ones discussed above, but combined applications of those discussed 
should allow for clear detection and characterization of the quantum phases discussed in this paper. 

Let us start this discussion with ideal Fermi gas, Fermi liquid and metaling phases between Fermi glass and 
glassy Mott insulator. All of them are fluids, i.e. will respond consequently to perturbations. They are gappless, 
and differ 
in this sense from the Fermi superfluids. All of them should lead to a nontrivial Fermi surface 
imaging in ballistic expansion. Difference between ideal, and interacting phases is here rather quantitative, 
and as such can be measured. Quantities such as the effective mass can be recovered from the measurements. 
Influence of disorded on these phase will be seen as gradual decrase of their "conducting" properties. 
Similar scenario  is expected to take place  with "dirty" superfluids; here measurements of the 
gap (using any of the excitation probing methods) should reveal rapid gap decrease with the increasing disorder. 

Disordered and glassy phases, such as Femi glass of glassy Mott insulator 
 are more difficult to detect. Obviously they will tend
 to give blurred images in the
measurement of the first order coherence. Spatial noise interferometry should reveal some information 
about the glassy Mott insulator, especially in the region of parameters where it will incorporate domains of 
checker board phase. Although the glassy phases  are gapless, 
the states forming 
the spectral quasi-continuum at low energies may be very difficult to achieve in simple excitation  measurements, 
since they may lie far away one from another in the phase space. Probing of one of such states  would thus allow 
to study excitations 
accesible  locally in the phase, which most presumably will be gapped. The character of excitations, 
and in particular their spectrum  should, however, be a
very sensitive function of how one excites them, and how one detects them (compare \cite{lukin}). 
On the other hand, domain insulator phase should be visible by "naked eye", and independent detection  
of bosons and composites. Also, the noise interferometry should reveal informations about the presence of the 
lattice, 
similarly as in the standard Mott insulator phase \cite{follig}. 

Finally, a separate problem concerns  detection of the fermionic spin glass phase and 
its properties, as well as distinction between the possible adequacy of Parisi versus droplet model. 
Repeated preparation of the system in the lattice with the same disorder 
(or even direct comparison and measurement of
overlap of replicas \cite{zollerhp})  will shed some 
light on the latter problem. In many other aspects, response of the spin glass to excitations will be similar 
to that of Fermi glasses and glassy Mott insulator.

%%%%%%%%%%%%%%%%%%%%%%%%%%%%%%%%%%%%%%%%%%%%%%%%%%%%%%%%%%%%%
\section{Conclusions}
\label{sec:conclusion}
Summarizing, we have studied atomic  Fermi-Bose mixtures in
optical lattices in the strong interacting limit, and in the
presence of an inhomogeneous, or random
 on-site potential. We have derived the effective Hamiltonian describing
the low temperature physics of the system, and shown that
an inhomogeneous potential may be efficiently
used to control the nature and strength of (boson mediated)
interactions in the system. Using  a  random potential, one is able to control
the system in such a way that its physics  corresponds to
a wide variety of quantum disordered systems. It is worth mentioning that the physics discussed in this paper
is very much analogous to the one of Bose-Bose mixtures in the limit of hard core bosons (when both species
exhibit strong intraspecies repulsion).

We end this section with a general comment on quantum complex systems. In our opinion quantum degenerate
gases offer an
absolutely unique possibility to study various models of physics of disorder systems, such as Bose
and Fermi glasses, quantum
percolating systems, "dirty" superfluids, domain and Mott insulators, quantum spin glasses,
systems exhibiting localization-delocalization phenomena, etc. The summary of predictions of this paper
is schematically shown in  the following list of quantum phases, obtained for
the case of only one type of sites, and $\mu_i^\textrm{f}=0$, or $\mu_i^\textrm{f}=\mu_i^\textrm{b}$:

\begin{itemize}
\item Composites $I$ - Fermi liquid, Fermi glass, quantum bond percolation;
\item Composites $II$ - ideal   Fermi gas,  Fermi glass, Fermi liquid, Mott Insulator, fermionic spin glass;
\item Composites $\overline{II}$ - domain insulator, "dirty" superfluid, Fermi glass, metallic phase,
Mott insulator, fermionic spin glass.
\end{itemize}
Additionally, for the case of lattices with different types of sites, physics of quantum site percolation
will become relevant.
Complex systems such as quantum celular automata, or neural networks can also be realized in this way.
In fact, we and other authors
have several times already stressed the fascinating possibility of using the ultracold lattice gases as
quantum simulators of complex systems. But, the proposed systems go beyond just repeating what is known from the
other kinds of physics; they allow to create novel quantum phases and novel quantum behaviors.

%%%%%%%%%%%%%%%%%%%%%%%%%%%%%%%%%%%%%%%%%%%%%%%%%%%%%%%%%%%%%
We thank  M. Baranov, H.-P. B\"uchler, B. Damski, M. Dressel, A. Georges, A. Niederberger, J. Parrondo, L. Santos, 
A. Sen (De), U. Sen, G.V. Shlyapnikov, J. Wehr and P. Zoller for fruitful discussions.
We acknowledge support from the Deutsche Forschungsgemeinschaft (SFB 407, SPP1116 436POL),
the RTN Cold Quantum Gases, ESF PESC QUDEDIS, the Alexander von Humboldt Foundation and Ministerio de Ciencia y Tecnologia (BFM-2002-02588). V.A. acknowledges support from the European Community (MEIF-CT-2003\
-501075).

\begin{appendix}
\section{Effective Hamiltonian to Second Order}\label{effhamil}
The Hamiltonian of our 2-species system described in
\ref{sec:Heffective} splits into two components: \HO\ and $H_{int}$. 
The Hamiltonian \HO\ has known eigenstates,
that are grouped in blocks (or manifolds)
of states close in energy while the 
differences between states from two different
blocks are much larger than the intrablock spacing.
In the case of section \ref{sec:Heffective}, this corresponds
to the manifold of near-ground states,
separated in energy by terms of order
$\Deltaij^\textrm{b,f}$, which is separated from other
blocks of higher excited states (two or
more bosons in one site) by terms of the
order \V. Generally, the projector on each
block space is denoted by \Palpha\, where \alph\ is the block index,
and the $i$-th state in any block is denoted
by \ketai. Note that 
$\Palpha\HO\Pbeta=0$ holds for $\alph\neq\beta$.

The second component is a Hamiltonian,
$\Hint$, that couples to \HO\ via a factor \J,
where \J\ is considered to be small, to form
the complete Hamiltonian of the system,
$H=\HO+J\Hint$. The interaction Hamiltonian
is now considered to introduce couplings
between block \alph\ and $\beta$, i.e.,
$\Palpha\Hint\Pbeta\neq0$ for $\alph\neq\beta$.

Following the technique detailed e.g. in \cite{cohen},
we construct an effective Hamiltonian,
\Heff\, from $H$ such that it describes the slow, low-energy
perturbation-induced tunneling {\it strictly within}
each manifold of unperturbed
block states, i.e., $\Palpha\Heff\Pbeta=0$ for
$\alph\neq\beta$, and has the same eigenvalues as $H$.
Tunneling between different blocks is thus neglected,
as this corresponds to fast, high-frequency processes
which we neglect here. Technically, the requirements
for \Heff\ are:
\begin{enumerate}
\item \Heff\ is hermitian, with the same
eigenvalues and the same degeneracies as $H$.
To achieve this, one defines \ensm{T:=e^{iS}},
with $S$ hermitian, $S=S^{\dagger}$, and chosen
such that:
\begin{equation}\Heff=THT^{\dagger} ~. \end{equation}
\item \Heff\ does not couple states from different
 manifolds:
\begin{equation}\Palpha\Heff\Pbeta=0,\quad\alph
\neq\beta\quad ~. \end{equation}
\item As the first two conditions still allow for
an infinite number of unitary transformations (all
$UT$ are still possible, $U$ being any unitary
transformation acting only {\it within} the manifolds),
the following additional condition is imposed:
\begin{equation}\Palpha S\Palpha=0\quad\mbox{for any}
\hspace{0.1cm}\alpha\end{equation}
\end{enumerate}
Expanding the first condition using the Baker-Hausdorff
formula, one obtains:
\begin{eqnarray}\lefteqn{\label{heff1}\Heff=H+\lsb
iS, H\rsb+\frac{1}{2!}\lsb iS, \lsb iS,
H\rsb\rsb}\hspace{1.0cm}\nonumber\\& &+\frac{1}{3!}\lsb iS, \lsb iS,
\lsb iS, H\rsb\rsb\rsb + \ldots
\end{eqnarray}
Making a power-series ansatz in $S$,
\begin{equation}\label{sexpand}S=J\Sone+
J^2\Stwo+J^3S_3+\ldots,\end{equation}
and employing $H=\HO+J\Hint$ one obtains
from (\ref{heff1}) to second order
\begin{eqnarray}
\Heff=\HO+
J\overbrace{\lrb\lsb iS_1, \HO\rsb+\Hint\rrb}
^{\Heff^1}\hspace{1.0cm}\nonumber\\
+J^2\overbrace{\lrb\lsb iS_2,
\HO\rsb+\lsb iS_1,\Hint\rsb+\frac{1}{2}\lsb
iS_1, \lsb iS_1, \HO\rsb\rsb\rrb}^{\Heff^2} ~.
\label{heff2}
\end{eqnarray}
This is a power series for \Heff, with its
moments denoted by \ensm{\Heff^1},
\ensm{\Heff^2}, $\ldots$ , where \ensm{\Heff^n}
generally depends on all \ensm{S_j} with \ensm{
1\leq j\leq n}. This allows for a systematic
evaluation of the matrix elements \ensm{\braai
iS_j\ketbj}, and consequently delivers matrix
expressions for the \ensm{\Heff^j} and \Heff.

To start with this, one considers the
expansions (\ref{heff2}) and (\ref{sexpand}) up
to first order, i.e. \ensm{\Heff=\HO+J\Heff^1}
and $S=J\Sone$. Using the second and third conditions,
as well as \ensm{\Palpha\HO\Pbeta=0} and the
expression for \ensm{\Heff^1} in (\ref{heff2}),
one finds:
\begin{equation}\label{sone1}\braai iS_1\ketbj
\lrb E_{\beta j}-E_{\alpha i}\rrb+\braai\Hint
\ketbj=0\end{equation}
\begin{equation}\label{sone2}
\Rightarrow\braai
iS_1\ketbj=\lgb \begin{array}{cr}\frac{\braai
\Hint\ketbj}{E_{\alph i}-E_{\beta j}}&\alph\neq
\beta\\0&\alph=\beta\end{array}\right. ~.
\end{equation}
Thus, the effective Hamiltonian within
the \alph -manifold, depends only on the
interaction term and not on \Sone, i.e., \ensm{
\braai\Heff^1\ketaj=\braai\Hint\ketaj}. 
A general result for any $n$ is that \ensm{\braai\Heff^n\ketaj} is independent of 
\ensm{S_n}. Based on the third condition, and on the observation that $S_n$ enters the
expression for \ensm{\Heff^n} only in the commutator
with \HO, which is diagonal in the manifold index.

Thus, when continuing to second order, the
term \ensm{\lsb iS_2, \HO\rsb} in the expression
for \ensm{\Heff^2} can be dropped. Of the two
remaining terms defining \ensm{\Heff^2} in
(\ref{heff2}), the second one can be simplified
by observing, that according to (\ref{sone1})
the operator \ensm{[iS_1, \HO]} is purely
non-diagonal in the manifold index, with values
opposite to those of the non-diagonal
part of the interaction
Hamiltonian. Thus, \ensm{\frac{1}{2}\lsb
iS_1, \lsb iS_1, \HO\rsb\rsb=-\frac{1}{2}\lsb
iS_1, \Hint^{nd}\rsb}. Now inserting the identity
between the operators in the still untreated
second term in \ensm{\Heff^2}, \ensm{\lsb i\Sone, \Hint\rsb} ,
one sees that due to \Sone\ being non-diagonal
in \alph, again only the non-diagonal part
of \Hint\ can contribute: \ensm{\lsb iS_1, \Hint\rsb
=\lsb iS_1, \Hint^{nd}\rsb}. Therefore, one has:
\begin{equation}\label{heff3}\Heff^2=[ iS_1,
\Hint^{nd}]+\frac{1}{2}[ iS_1, \overbrace{
\lsb iS_1, \HO\rsb}^{-\Hint^{nd}}]=\frac{1}{2}
[iS_1, \Hint^{nd}] ~.
\end{equation}
Collecting all terms relevant for \braai\Heff
\ketaj\ to second order in \J, and introducing
the notation \ensm{\Qalphai:=\sum_{k,\gamma\neq
\alph}\frac{|\gamma,k\rangle\langle\gamma,k|}
{E_{\gamma k}-E_{\alpha i}}}, one finds:

\begin{eqnarray}\label{heff4}\braai\Heff\ketaj
=E_{\alph i}\delta_{ij}+J\braai\Hint\ketaj \hspace{0.0cm}\nonumber\\
-\frac{J^2}{2}\lrb\braai\Hint\lsb\Qalphai+
\Qalphaj\rsb\Hint\ketaj\rrb
\end{eqnarray}
where the identity operator has been inserted
in the final expression for \ensm{\Heff^2} in
formula (\ref{heff3}), and then evaluated using
formula (\ref{sone2}), which naturally leads one
to define the operator \Qalphai\ as above.
Note that this construction can be generalized to
arbitrary orders in \J\ in a straightforward
manner, as detailed in \cite{cohen}.

\section{Stability of the SK-solution}
\label{atsubsec}
The Taylor-expansion of Eq.~(\ref{sk6a}) around $\qab=q$ yields the correction \ensm{-\frac{1}{2}\delta\overline{f}}
to the free energy, with:
\begin{widetext}
%++++++++++++++++++++++++++++++%
\begin{equation}\label{at1}\delta\overline{f}=\sum_{\lsb\alph,\beta\rsb\lsb\gamma,\delta\rsb}
\underbrace{\left.\frac{\partial^2\overline{f}}{\partial\qab\partial
q_{\gamma\delta}}\right|_{SK}}_{:=G_{(\alph\beta)(\gamma\delta)}}\dqab\dqcd=
\sum_{\lsb\alph,\beta\rsb\lsb\gamma,\delta\rsb}
\lsb\delta_{(\alph,\beta)(\gamma,\delta)}-(\beta K)^2
\lrb\langle\sa\sbb\scc\sd\rangle^{SK}_L-\langle\sa\sbb\rangle^{SK}_L
\langle\scc\sd\rangle^{SK}_L\rrb\rsb\dqab\dqcd
\end{equation}
%++++++++++++++++++++++++++++++%
\end{widetext}
where $\lsb\cdot,\cdot\rsb$ denotes a sum over all distinct index pairs irrespective of the order,
the index $SK$ denotes evaluation at $\qab=q$ and, analogous to (\ref{sk11}) and (\ref{sk12}), \ensm{\left.\langle\sa\sbb
\scc\sd\rangle_L\right|_{SK}=\lrb\frac{1}{2\pi}\rrb^{\frac{1}{2}}\infint dz
e^{-\frac{z^2}{2}}\tanh^4(\Az)}.

If the SK solution really corresponds to a maximum, this symmetric quadratic form
must be positive definite. To check this, one calculates
the eigenvalues of the \ensm{\frac{1}{2}n(n-1)}-dimensional
matrix \ensm{G_{(\alph\beta)(\gamma\delta)}} in terms of its
three distinct matrix elements:
%++++++++++++++++++++++++++++++%
\begin{eqnarray*}
P&\equiv&G_{(\alph\beta)(\alph\beta)}=1-\left.(\beta K)^2(1-\langle\sa\sbb\rangle_L^2)\right|_{SK}\\
Q&\equiv&G_{(\alph\beta)(\alph\gamma)}=
-(\beta K)^2\left.(\langle\sbb\scc\rangle_L-\langle\sa\sbb\rangle_L^2)\right|_{SK}\\
R&\equiv&G_{(\alph\beta)(\gamma\delta)}=
-(\beta K)^2\left.(\langle\sa\sbb\scc\sd\rangle_L-\langle\sa\sbb\rangle_L^2)\right|_{SK} ~.\\
\end{eqnarray*}
%++++++++++++++++++++++++++++++%
As becomes apparent from this, there are just three distinct classes of transformations that leave \ensm{G_{(\alph\beta)(\gamma\delta)}}
invariant: those that permute no index acting on the $P$'s; those that permute a single index acting on the $Q$'s
; and those that permute two indices acting on the $R$'s.
Thus, there are only three eigenspaces to distinct eigenvalues and the linearly independent eigenvectors within each of these
eigenspaces can naturally be chosen by considering a group of vectors that is invariant under the corresponding permutation transformation.

\begin{figure}[t!]
\mbox{\epsfxsize 3.00in \epsfbox{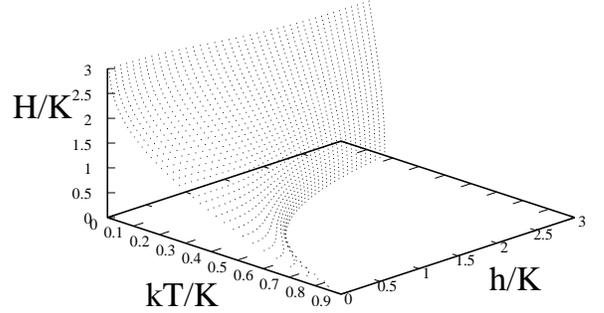}}
\caption{Almeida-Thouless plane in the reduced variables $kT/K$, $h/K$ and
$H/K$. The constraint on the magnetization (\ref{sk12}) has been
effectively neglected for this, as it is an parameter in the experiment
as well.}
\label{atplane}
\end{figure}
{\it 1. No index permutation:}

%++++++++++++++++++++++++++++++%
The ansatz for the eigenvalue is trivially given by:
\begin{equation}\label{ev1}
\delta\qab=c,\hspace{0.3cm}\mbox{for all $(\alpha\beta)$},
\end{equation}
%++++++++++++++++++++++++++++++%
which is nondegenerate.
With this ansatz, the eigenvalue equation is
%++++++++++++++++++++++++++++++%
\begin{equation}
\label{atv0}\lsb P+2(n-2)Q+\frac{1}{2}(n-2)(n-3)R-\lambda_1\rsb c=0
\end{equation}
%++++++++++++++++++++++++++++++%
from which $\lambda_1$ can immediately be read off.

{\it 2. Permutation of a single index:}

Again, the ansatz for the eigenvectors is naturally given by:
%++++++++++++++++++++++++++++++%
\begin{equation}\label{ev2}
\delta\qab=c,\quad\mbox{for $\alpha$ or $\beta=\theta$},\hspace{0.3cm}\delta\qab=d,\quad\mbox{for $\alpha$, $\beta\neq\theta$} ~.
\end{equation}
%++++++++++++++++++++++++++++++%
The ansatz still contains the previous case, as the requirement of the eigenvectors (\ref{ev1}) and (\ref{ev2})
being orthogonal still has to be fulfilled. This yields $c=(1-n/2)d$ and a degeneracy of $n-1$ for the eigenvalue.
The eigenvalue equation then becomes:
%++++++++++++++++++++++++++++++%
\begin{equation}
\label{atv1}\lsb P+(n-4)Q+(n-3)R-\lambda_1'\rsb d=0
\end{equation}
%++++++++++++++++++++++++++++++%
from which $\lambda_1'$ is again immediately obvious.

{\it 3. Permutation of both indices:}

The ansatz is:
%++++++++++++++++++++++++++++++%
\begin{equation}\label{ev3}\begin{array}{c}\delta q_{\theta\nu}=c,\hspace{0.3cm}\delta q_{\theta\alpha}=\delta q_{\nu\alpha}=d\hspace{0.3cm}\mbox{for $\alpha\neq\theta$, $\nu$}
\\ \delta\qab=e\hspace{0.3cm}\mbox{for $\alpha$, $\beta\neq\theta$, $\nu$}\end{array}
\end{equation}
%++++++++++++++++++++++++++++++%
Orthogonality to the previous eigenspaces requires $c=(2-n)$, $d=(3-n)/2$
and results in the final eigenvalue equation:
%++++++++++++++++++++++++++++++%
\begin{equation}
\label{atv2}\lsb P-2Q+R-\lambda_2\rsb e=0
\end{equation}
%++++++++++++++++++++++++++++++%
with $\lambda_2$ having degeneracy \ensm{\frac{1}{2}n(n-3)}.

Taking the naive replica limit again the first two eigenvalues coincide:
\begin{equation}
\lim_{n\rightarrow 0}\lambda_1=\lim_{n\rightarrow 0}\lambda_1'=P-4Q-3R ~. \\
\end{equation}
As de Almeida and Thouless report \cite{at}, a region in parameter-space where
this limiting value took negative value could not be found. Thus,
calculating $\lambda_2$ explicitly, the relevant stability condition reads:
%++++++++++++++++++++++%
\begin{equation}\label{at5}
\lambda_2=\frac{1}{(\beta K)^2}-\frac{1}{\sqrt{2\pi}}\infint dz e^{-\frac{z^2}{2}}\sech^4(\Az)>0 ~.
\end{equation}
%++++++++++++++++++++++%
Solving the coupled equations (\ref{sk10}) - (\ref{sk12}) for
$\lambda_2=0$ yields the dAT-plane (cf. figure
\ref{atplane}). Above it, the SK-solution is still valid, and below it
$\lambda_2$ takes negative value and the SK-solution breaks down.

This can of course only happen if eigenvectors to $\lambda_2$ are compatible
with the magnetization constraint (\ref{sk7a}).
To see this, one has to check that small fluctuations around $\qab=q$ do not lead to a
deviation from the value for $m$, i.e.,that:
%++++++++++++++++++++++%
\begin{equation}\label{magconstraint}
\delta m=\sum_{\lsb\alpha,\beta\rsb}\left.\frac{\partial\langle s^{\lambda}\rangle_L}{\partial\qab}\right|_{SK}
\delta\qab=0
\end{equation}
%++++++++++++++++++++++%
holds, with:
%++++++++++++++++++++++%
\begin{eqnarray}
\left.\frac{\partial\langle s^{\lambda}\rangle_L}{\partial\qab}\right|_{SK}= \hspace{4.4cm}\label{magderiv}\\
\lgb \begin{array}{ll}\left.\lrb(K\beta)^2\langle s^{\lambda}\rangle_L -\langle\sa\sbb\rangle_L\langle s^{\lambda}\rangle_L\rrb\right|_{SK}&\hspace{0.0cm}\mbox{$\alpha$ or $\beta=\lambda$} \nonumber\\
\left.\lrb(K\beta)^2\langle s^{\lambda}\sa\sbb \rangle_L -\langle\sa\sbb\rangle_L\langle s^{\lambda}\rangle_L\rrb\right|_{SK}&\hspace{0.0cm}\mbox{$\alpha$, $\beta\neq\lambda$}\end{array}\right. ~.
\end{eqnarray}
%++++++++++++++++++++++%
Inserting any eigenvector of $\lambda_2$ into (\ref{magconstraint}) will however yield the desired result.
As is clearly seen from (\ref{magderiv}) $\partial\langle s^{\lambda}\rangle_L/\partial\qab$ evaluated at $\qab=q$  is an eigenvector to
$\lambda_1'$, and thus $\delta m=0$ is fullfilled.

The magnetization constraint is thereby compatible with the instability $\lambda_2<0$,
and replica-symmetry breaking is expected to occur.
%The Parisi-M\'ezard approach was invented to provide a stable solution
%below the AT-plane. It consists of an ansatz for \qab\ in (\ref{sk7})
%that is different from the SK one (cf. the reprints 8-11 and 16-17
%in \cite{mez87} for details) in that it breaks replica-symmetry, i.e.
%\ensm{\qab=\langle\sia\sib\rangle_L} now does depend on the chosen
%replica-indices. Considering the unambiguity in equation (\ref{rsb7})
%regarding the indices, it becomes clear that in this mean-field formulation
%\qab\ can no longer denote the disorder-averaged overlap parameter. Instead
%a sensible redefinition consists of an average over all different
%solutions, i.e.
%\begin{equation}q=\label{at7}\lim_{n\rightarrow0}\frac{1}{n(n-1)}\sumab\qab
%\end{equation}
%Therefore, the averaged overlap distribution in mean-field theory is
%thus simply given by
%\ensm{P(q)=\lim_{n\rightarrow0}\frac{1}{n(n-1)}\sumab\delta(q-\qab)}
%and which is the mean-field equivalent of $\PQ$ from \ref{sksubsec}.
%
%It is this mean-field RSB-theory that underlies the theoretical proposal
%for the EA-model that was briefly sketched  in \ref{sksubsec}.
>From the calculations it has become clear that the presence of the random magnetic
field would not change the occurrence of {replica symmetry breaking}, and all properties like
infinitely degenerate ground-states and ultrametricity would be expected to
occur in this model as well, provided the M\'ezard-Parisi approach can be
applied to the finite range spin-glass at all.

\end{appendix}

%%%%%%%%%%%%%%%%%%%%%%%%%%%%%%%%%%%%%%%%%%%%%%%%%%%%%%%%%%%%%


\begin{references}
\bibitem{foot2}In all the work presented below, we considered a
sufficiently large $n_\textrm{max}$ so that the numerical results are not affected by its finite value.
\bibitem{foot3} Note that we do not use standard grand
canonical ensemble corresponding to
$\langle H_\textrm{FBH} \rangle -\mu^\textrm{f} \langle \sum_i m_i -N_\textrm{f} \rangle -
\mu^\textrm{b} \langle \sum_i n_i -N_\textrm{b} \rangle$ because in the case of composites,
we checked that it turns out to be very difficult to choose the chemical
potentials $\mu^\textrm{f}$ and $\mu^\textrm{b}$ in order to get reasonable
values of $\langle \sum_i m_i \rangle$ and $\langle \sum_i n_i \rangle$. This is because
both total numbers of particles depend on both $\mu^\textrm{f}$ and $\mu^\textrm{b}$ in a
non straightforward way. On the contrary, using $\Lambda_\textrm{f}$ and $\Lambda_\textrm{b}$
large enough and positive ensures that $\langle \sum_i m_i \rangle \simeq N_\textrm{f}$ and
$\langle \sum_i n_i \rangle \simeq N_\textrm{b}$ without effort.
\bibitem{foot4}Although not presented in this paper, we also performed calculations in the regions corresponding
to composites of types $I$ and $II$ (see composites diagram in Fig.~\ref{fig:phase}) that demonstrate formation
of the corresponding composites. For instance, for type $I$ composites, $N_\textrm{b}=100$ and $N_\textrm{f}=40$,
 we found that the only non-zero probabilities were $|g^{(i)}_{n=1,m=0}|^2$ (absence of a composite) and
$|g^{(i)}_{n=1,m=1}|^2$ (presence of a composite) with $|g^{(i)}_{n=1,m=0}|^2+|g^{(i)}_{n=1,m=1}|^2=1$.
Similarly, for type $II$ composites, $N_\textrm{b}=140$ and $N_\textrm{f}=40$, we found that the
only non-zero probabilities were $|g^{(i)}_{n=1,m=0}|^2$ (absence of composite) and $|g^{(i)}_{n=2,m=1}|^2$
(presence of composite) with $|g^{(i)}_{n=1,m=0}|^2+|g^{(i)}_{n=2,m=1}|^2=1$.
\bibitem{foot5} {As the standard deviation of \Kij\
is given by $K/\sqrt{N}$, it vanishes in the thermodynamic limit. This choice is necessary on the
technical level to obtain a sensible limit. Physically this corresponds to the case where the impurities are
very dense (each bond is a separately fluctuating impurity) and influence the system collectively,
but are negligible individually.}
\bibitem{foot6} {Generally, the lower critical dimension for
this transition to happen at non-zero $T$ is believed to be two (cf. references in \cite{stein})}.
\bibitem{foot7}{In an ultrametric space,
the distance measure does not obey the triangle inequality,
but the stronger relation $d(x,y)\leq\max(d(x,z),d(y,z))$. Pictorially,
the distances between any three points $x$, $y$ and $z$ can only form an equilateral or an
acute isosceles triangle.}
\bibitem{zollerhp} H. -P. B\"uchler and P. Zoller proposed to prepare arrays of 2D lattices with randomly distributed
transverse columns of chains of auxiliary atoms, that could serve as the same distribution of random scatterers
in each  2D lattice.
\end{references}
\end{document}